\documentclass[aps,pra,english,floatfix,showpacs,twocolumn]{revtex4}
\usepackage[english]{babel}
\usepackage[ansinew]{inputenc}
\usepackage{amsfonts}
\usepackage{latexsym}
\usepackage{amsmath}
\usepackage{amssymb}
\usepackage{graphicx}
\usepackage{dcolumn}

\begin{document}

\title{Mixing of ultracold atomic clouds by merging of two magnetic traps}
\author{Jesper Fevre Bertelsen}
\author{Henrik Kjaer Andersen}
\author{Sune Mai}
\author{Michael Budde}
\affiliation{Danish National Research Foundation Center for Quantum Optics\\Department of
Physics and Astronomy, University of Aarhus\\DK-8000 Aarhus C, Denmark}

\date{\today}

\begin{abstract}
We demonstrate a method to make mixtures of ultracold atoms that does not make use of a two-species magneto-optical trap. We prepare two clouds of $^{87}$Rb atoms in distinct magnetic quadrupole traps and mix the two clouds by merging the traps. For correctly chosen parameters the mixing can be done essentially without loss of atoms and with only minor heating. The basic features of the process can be accounted for by a classical simulation of particle trajectories. Such calculations indicate that mixing of different mass species is also feasible, opening the way for using the method as a starting point for making quantum gas mixtures.
\end{abstract}

\maketitle

\section{Introduction}
Production of ultracold atomic gases consisting of a mixture of two different elements is interesting for several reasons. First of all it makes it possible experimentally to study and exploit inter-species interactions \cite{InguscioRbK,JinRbK,Bongs,Zimmermann,Sympathetic8587,KetterleNaLi,Truscott,Salomon,CsLi,NaCs} including inter-species Feshbach resonances \cite{InguscioFeshbach,NaLiFeshbach,JinFeshbach,Scott}. An important possibility in this respect is to produce ultracold heteronuclear molecules with strong electric dipole-dipole interactions using photoassociation \cite{StwalleyMolecules,BigelowMolecules,deMilleMolecules,WeidemullerMolecules,ManciniMolecules} or Feshbach resonances \cite{BongsMol,Scott,JinMol,KokkelmansFeshbach,DevreeseFeshbach}. Another is the study of phase separation phenomena \cite{InguscioPhaseSep,Bongs,Molmer,Pethick} and realization of fermionic superfluidity mediated by bosons \cite{Viverit,Bijlsma,Viverit2}. Furthermore mixtures in optical lattices is a new and relatively unexplored field of study with many interesting phenomena \cite{BongsMol,BongsLattice,EsslingerLattice,Lewenstein,LewensteinReview,StoofLattice,Damski}.

The mixtures also have an important application in that one of the species can be used to sympathetically cool the other one. If one of the species is evaporatively cooled then the other species - which might have scattering properties which make evaporative cooling impossible - will be cooled too if the inter-species thermalization rate is large enough. This principle has been used to cool fermionic atoms to quantum degeneracy \cite{Truscott,Salomon,KetterleNaLi,InguscioRbK,JinRbK,Bongs,Zimmermann} and can also be used to condense bosonic isotopes which are otherwise difficult to condense as demonstrated for $^{41}$K in Ref. \cite{InguscioK}.

Currently many groups have produced ultracold mixtures in the quantum regime. This includes boson-fermion mixtures where the fermions are cooled sympathetically by the bosons ($^7$Li/$^6$Li \cite{Truscott,Salomon}, $^{23}$Na/$^6$Li \cite{KetterleNaLi}, $^{87}$Rb/$^{40}$K \cite{InguscioRbK,JinRbK,Bongs}, $^{87}$Rb/$^6$Li \cite{Zimmermann}) and very recently $^{3}$He/$^{4}$He \cite{HeHe} and boson-boson mixtures ($^{87}$Rb/$^{41}$K \cite{InguscioK}, $^{87}$Rb/$^{85}$Rb \cite{Scott,Sympathetic8587}). Furthermore several other mixtures in the ultracold, but not yet quantum, regime have been studied (e.g. \cite{CsLi,NaCs,BigelowMolecules,deMilleMolecules,ManciniMolecules,WeidemullerMolecules,Cr}).

Even if the goal is to enter the quantum regime by means of evaporative cooling the starting point is a mixture which has already been cooled to the 100 $\mu$K regime by means of laser cooling. The standard approach used to produce such cold mixtures is to cool and trap the two (or even three \cite{3MOT}) species simultaneously in the same magneto-optical trap (MOT).

In this article we demonstrate a completely new method for making the mixtures. The basic idea is to cool and trap the two elements in separate magneto-optical traps and mix the two clouds afterwards. After having cooled the atoms we trap the two clouds in distinct magnetic quadrupole traps and mix them by merging the two traps mechanically. Since the magnetic fields from the two traps interfere during the merging process it is not trivial that one can mix atomic clouds in this way. We have demonstrated that the method is indeed feasible and a promising new way of producing mixtures of cold atoms, but it needs accurate control of various parameters.

The procedure makes it possible to optimize the cooling and trapping process of each species individually and it is a modular approach where one can build the setup needed for producing one species and then add the other one when needed. Furthermore there is no need for optics which can handle two different wavelengths.

In section \ref{ExpSetup} we give an overview over our experimental setup. Section \ref{Field} is devoted to a discussion of the trapping potential and its evolution during the mixing process. In section \ref{Model} we introduce a simple theoretical model which we use to simulate the atomic particle trajectories in order to make predictions about the outcome and understand the physics of the mixing process. In section \ref{Results} we discuss our experimental results and compare them with the results of the theoretical model which turn out to reproduce the essential features quite well. We examine the dependence on essential parameters such as relative and absolute trap depths. Finally in section \ref{Outlook} we discuss mixing of different elements based on the theoretical simulations and conclude in section \ref{Conclusion}.

\section{Experimental Setup}\label{ExpSetup}
\begin{figure}[!]
\includegraphics[width=\columnwidth]{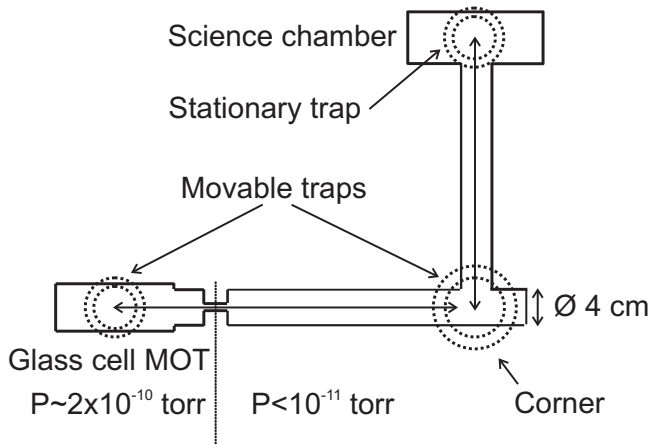}
\caption{\small Sketch of our vacuum chamber. The coils constituting the magnetic traps are symbolized by the dashed circles. The distance from the MOT to the corner and from the corner to the science chamber are 49 cm and 37 cm, respectively.}
\label{setup}
\end{figure}

Our experimental setup is sketched in Fig. \ref{setup}. The vacuum chamber consists of two parts separated by a differential pumping hole: a "high pressure" part ($P\sim 2\times 10^{-10}$ torr) with a cylindrical glass cell where we cool and trap $^{87}$Rb atoms in a MOT \cite{Lewandowski} and a "low pressure part" ($P<10^{-11}$ torr) where the lifetime due to rest gas collisions is about 2 min.

We have two magnetic quadrupole traps which are mounted on computer controlled mechanical positioning systems like in Ref. \cite{Lewandowski}. One trap is for moving the atoms from the MOT to the corner of the chamber and the other can move them from the corner to the science chamber (see the sketch in Fig. \ref{setup}). The science chamber is made of stainless steel and has its own magnetic  quadrupole trap which can be turned off in 340 $\mu$s allowing us to measure temperatures by time-of-flight imaging. The science chamber is furthermore used to make Bose-Einstein condensates (BECs) by standard forced radio frequency evaporation in a Quadrupole-Ioffe configuration (QUIC) trap \cite{QUIC}.

The setup is prepared for the addition of a second source chamber at the right side of the corner and this part will have its own movable magnetic trap. The idea is to mix the two elements at the corner in the future.

To study the principle of mixing by merging two magnetic traps we mix two clouds of $^{87}$Rb in the $|F=1,\ m_\textrm F=-1>$ hyperfine state. To produce the cooling light for the MOT we use a standard diode laser setup \cite{Lewandowski,MacAdam,Hollberg}. During the last 5 ms of the MOT phase we compress the atoms in the MOT by detuning the cooling lasers and decreasing the power in the repump to minimize heating when the magnetic trap is turned on \cite{Lewandowski}. The trap coils provide the magnetic field of the MOT. After the compressed MOT phase about 40 \% of the atoms are caught in the magnetic trap by ramping the current in the coils up to 150 A during 1 ms. Afterwards we wait for 55 ms and increase the current to 250 A in 200 ms before moving the atoms through the differential pumping hole.

\begin{figure}[!]
\includegraphics[width=\columnwidth]{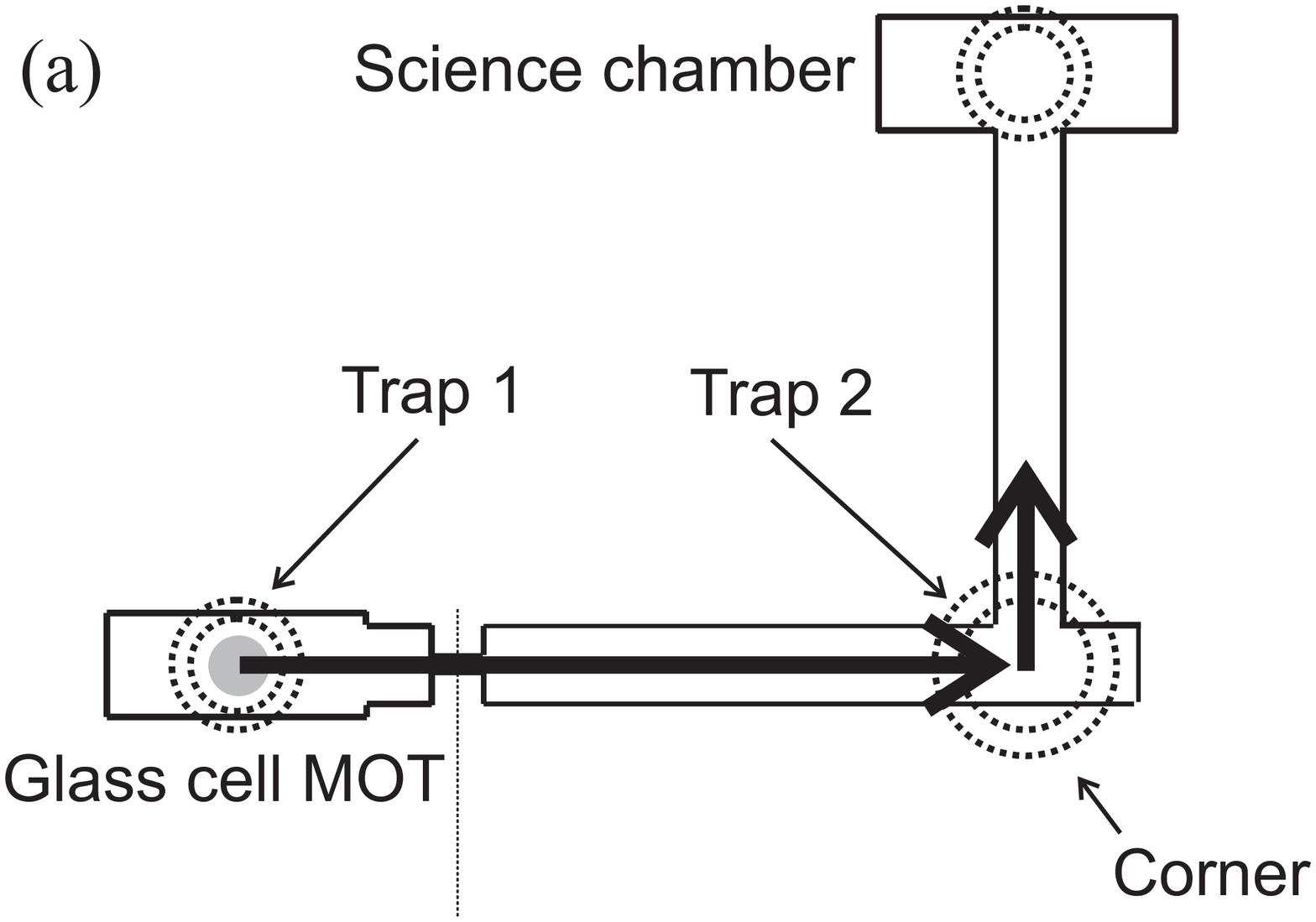}
\\
\includegraphics[width=\columnwidth]{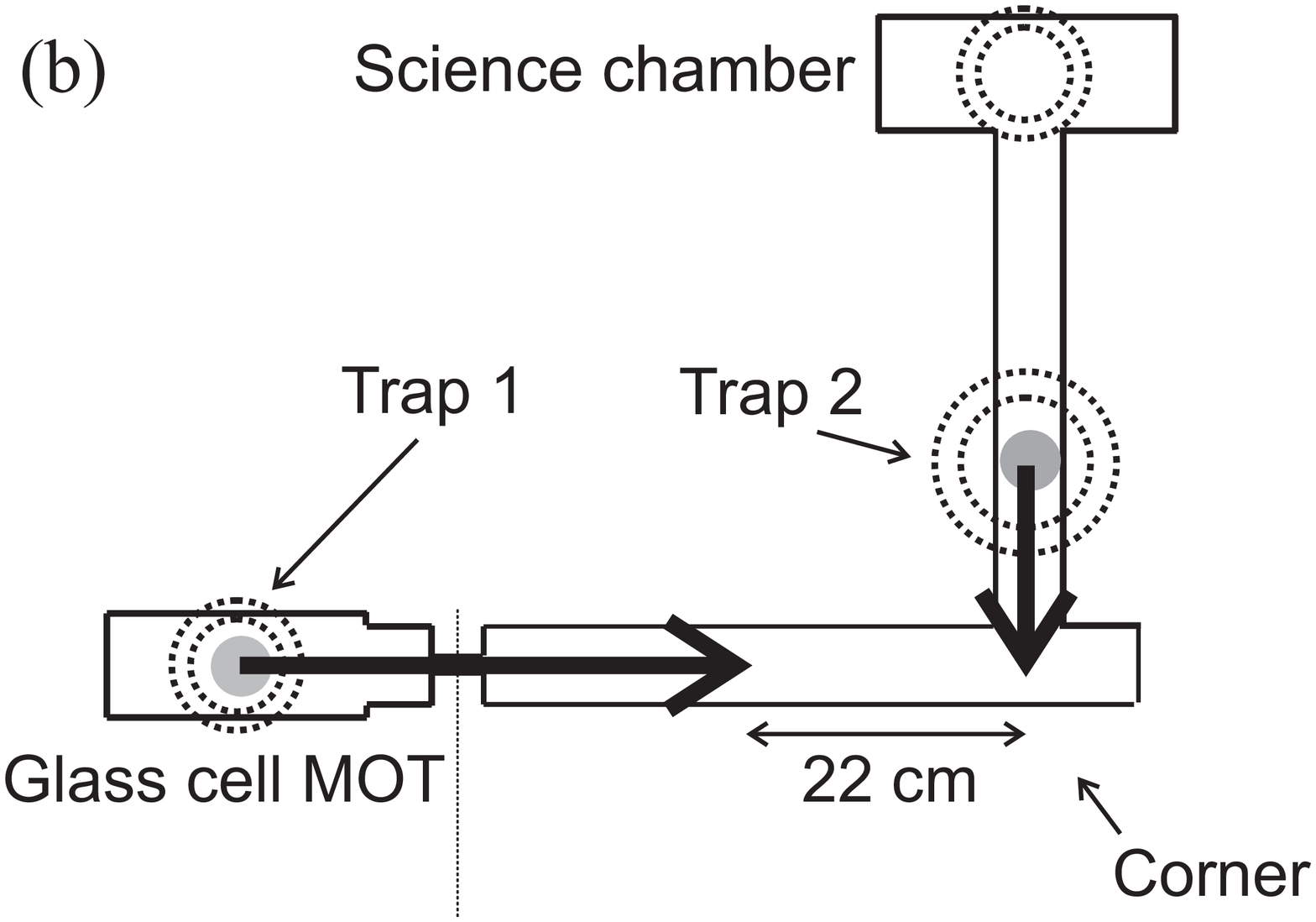}
\\
\includegraphics[width=\columnwidth]{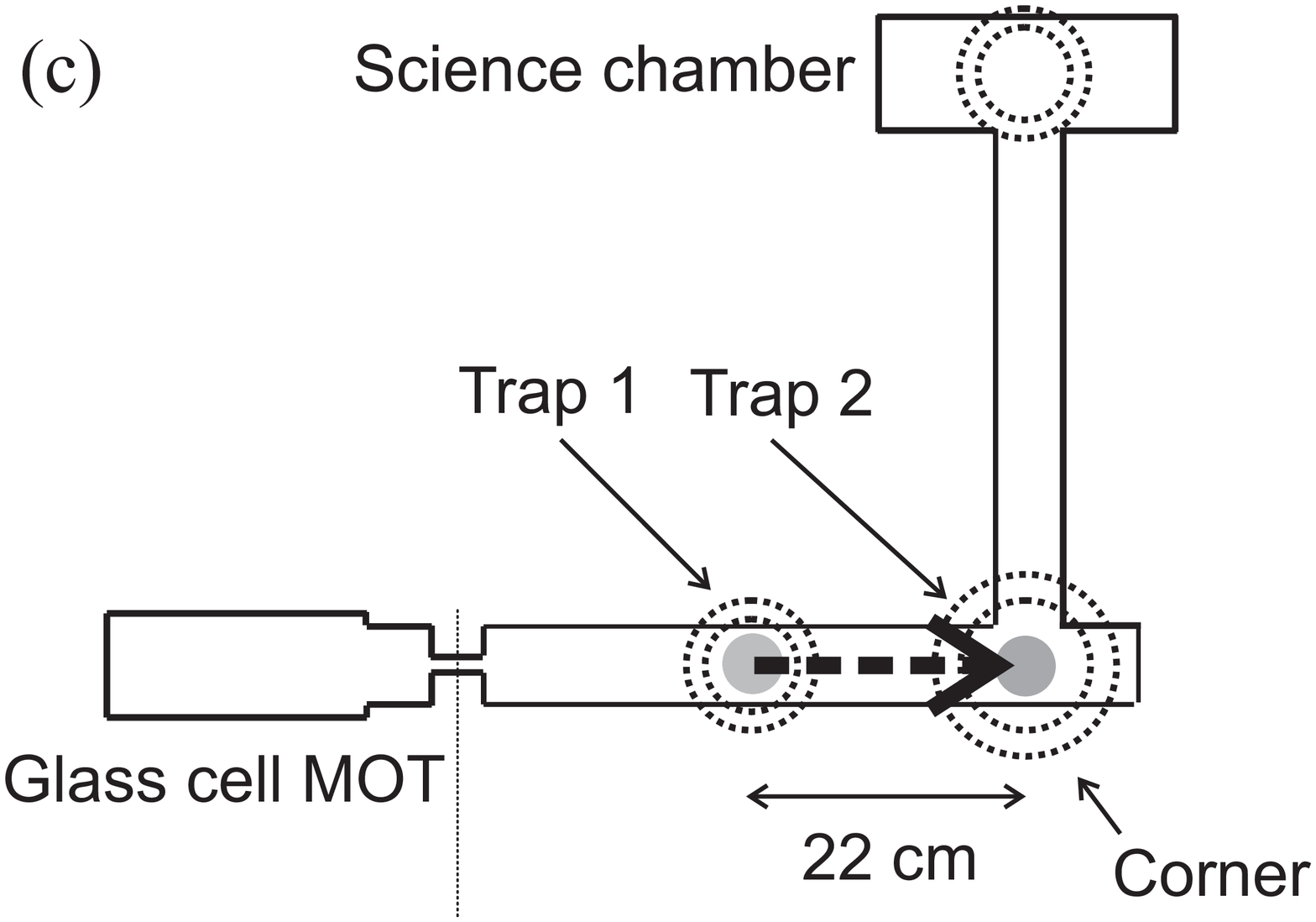}
\caption{\small The procedure used to mix two Rb clouds. (a): First we collect atoms in the MOT, trap them in Trap 1 and move the trap to the corner where the atoms are transferred into the other movable trap (Trap 2). This trap is moved 7 cm towards the science chamber to shield the atoms from scattered light. (b): Trap 1 is moved back and a second cloud of atoms is collected. This cloud is moved into the "low pressure" part of the chamber to a position 22 cm from the corner. The trap current is slowly changed to the initial value wanted during the mixing. Trap 2 is moved back to the corner. (c): Finally we merge the two traps by slowly (5 cm/s) moving Trap 1 towards the corner until the trap axes are 2 cm apart. This is the mixing process. We cannot move the traps closer together due to the limited space around the vacuum chamber. For the same reason Trap 2 is displaced 5 mm from the transport axis of Trap 1.}
\label{procedure}
\end{figure}

The procedure which we use to mix the two atomic clouds is explained in detail in Fig. \ref{procedure}. When we have caught atoms in the traps, we move Trap 1 towards Trap 2 at a speed of 5 cm/s. Trap 2 is stationary during the process. Both clouds are situated in the "low pressure" part of the chamber during the process. After the mixing process we move the atoms to the science chamber and transfer the atoms to the science chamber quadrupole trap. Here we wait 10 s to allow for thermalization before we take absorption images after 7, 9, 11 and 13 ms time-of-flight to measure the temperature and number of atoms.

In order to study how each of the atomic clouds is influenced by the process we can choose to have atoms in one or the other of the two traps or in both of them. We do this by having the cooling light either on or off during each of the two MOT phases. 

\section{Magnetic field and trapping potential}\label{Field}
Rb atoms can be trapped using the linear Zeeman shift \cite{QuadraticCorrection}. For a magnetic field of magnitude $B(\mathbf r)$ the trapping potential is $U(\mathbf r)=g_\textrm{F}m_\textrm{F}\mu_B B(\mathbf r)$ where $g_\textrm{F}=-1/2$ for the $F=1$ hyperfine state and $g_\textrm{F}=1/2$ for the F=2 hyperfine state of the ground term.

Our magnetic traps are simple quadrupole traps ideally consisting of two circular current loops with radius R and separation distance 2A. Around the zero-point in the middle of the trap the magnetic field increases linearly in all directions with the axial gradient being twice the radial gradient. Situating the origin in the middle we will refer to the axis through the coil centers as the z axis, and the radial coordinate will be called $\rho$. Ref. \cite{Bergeman} gives analytical expressions containing elliptic integrals for the magnetic field from circular current loops.

Near the minimum (for our coils up to about 1 cm from the minimum) the magnitude of the magnetic field can be well approximated by \cite{MetcalfStraten}
\begin{align}\label{fieldapprox}
B(\rho,z)\approx\frac{\partial |B|}{\partial\rho}\sqrt{\rho^2+4z^2}
\end{align}
where $\partial |B|/\partial\rho$ is the radial magnetic field gradient. This formula shows that the potential energy isosurfaces near the minimum are ellipsoids and it makes it possible to make analytical calculations of thermodynamic quantities (section \ref{Thermodynamics}).

The coils of Trap 1 and Trap 2 consist of, respectively, 16 and 31 windings of 4.25$\times$4.25 mm$^2$ square, copper tubing with cooling water flowing in the middle. The coils are cast in epoxy for stability and each coil has outer dimensions of about 10$\times$10$\times$2 cm.

We have calculated the magnetic field from our coils using a numerical implementation of the Biot-Savart law. As it turns out these calculations show that for our purposes the magnetic field from the coils can be very accurately approximated by the field from two circular windings, if $A$, $R$ and $I$ are chosen properly. The validity of this approximation makes the numerical simulations much faster and shows that if the mixing process works with our coils, it should work for other quadrupole traps, too.

Due to the interference between the fields from the two traps during the mixing process, the magnetic field from each trap needs to be accurately determined in the radial direction even at distances up to 20 cm. Therefore, to find optimum values of $A$, $R$ and $I$, we fit the Biot-Savart calculation for $B_\rho(\rho)$ in the $z=0$ plane with 1 A through the actual coils with the corresponding function $B_\textrm{A,R,I}(\rho)$ for a single pair of circular windings using $A$, $R$ and $I$ as fitting parameters. This representation is very good: Within 2 cm from the z=0 plane the maximum deviation is 7 mG and this occurs in the middle of the trap where the numerical Biot-Savart implementation is inaccurate due to the discontinuity of the field gradient. The best fitting values of $A$, $R$ and $I$ are listed in Table \ref{coilpar}.

\begin{table}[!htb]
\caption{Circular coil pair parameters.}
\begin{ruledtabular}
\begin{tabular}{l|dd}

Coil & \textrm{Trap 1} & \textrm{Trap 2}
\\
\hline
A (cm) & 3.941 & 7.082
\\
R (cm) & 3.2 & 4.21
\\
I (A) & 16.0 & 32.17
\\
Radial field gradient (G/cm)/A & 0.36 & 0.20
\\
Field barrier height (G/A) & 0.77 & 0.59
\end{tabular}
\end{ruledtabular}
\label{coilpar}
\end{table}
The maximum current we can use in our coils is 400 A which is limited by the current supply.

\begin{figure}[!htb]
\includegraphics[width=\columnwidth]{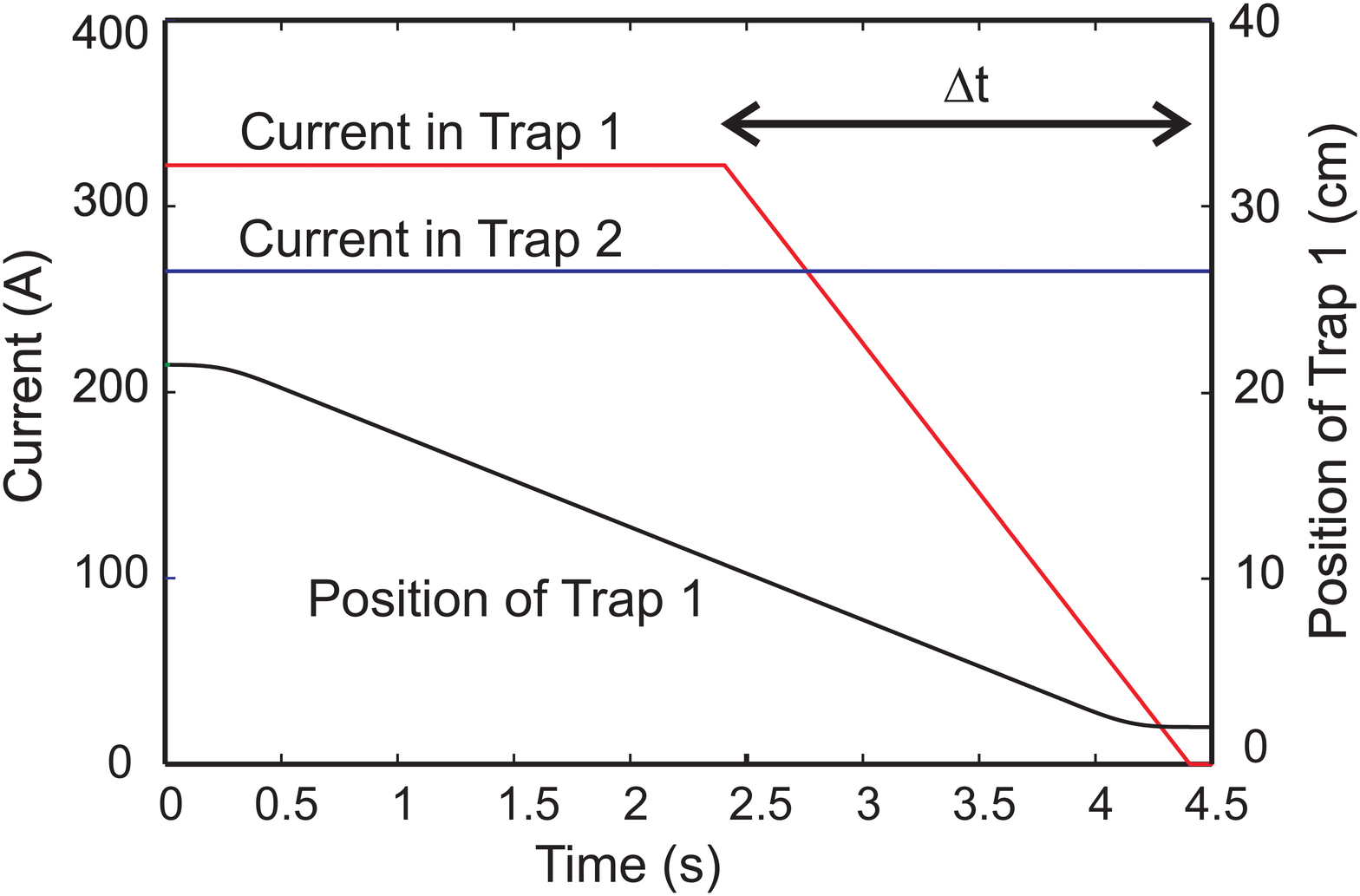}
\caption{\small (Color online) Current ramps and the position of Trap 1 during the mixing process.}
\label{currents}
\end{figure}

We use the current ramps sketched in Fig. \ref{currents}. The current in the coils constituting Trap 2 is constant throughout the mixing process. The current in the moving Trap 1 is constant in the beginning but is linearly ramped down to 0 towards the end such that it is 0 exactly when the movement stops. Also shown is the position of Trap 1 as a function of time. We use soft acceleration and deceleration phases (with constant time derivative of the acceleration), but during most of the time the velocity is constant.

We have tried to vary the time $\Delta t$ during which we ramp down the current of Trap 1 keeping the total mixing time constant (see Fig. \ref{currents}). We have found that the exact value of $\Delta t$ is not critical as long as it is larger than 1.5 s. Ramping down the current of Trap 1 faster than that causes additional heating (but essentially no loss of atoms). The optimal ratio of initial currents in the two traps is strongly dependent on the value of $\Delta t$.

We have also tried to make deviations from the current ramp in Fig. \ref{currents} by ramping down in two linear steps instead of one. This, however does not lead to any improvement of the results. One could imagine that a suitably tailored soft curve would lead to marginally better results, but since the results using the simple linear ramp are quite satisfactory we have not pursued this issue further.

\begin{figure}[!]
\includegraphics[width=\columnwidth]{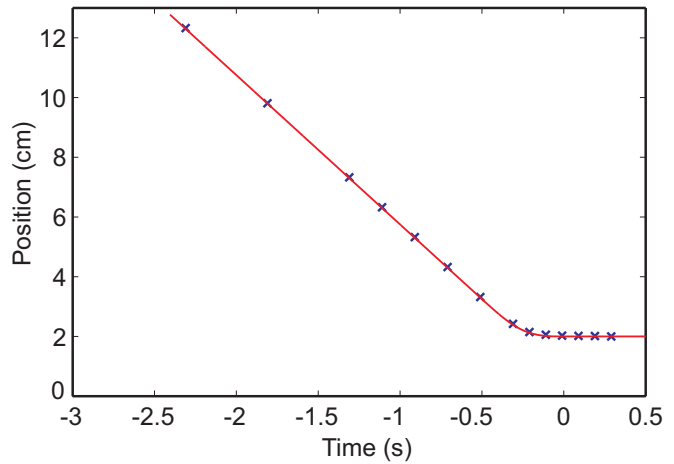}
\caption{\small (Color online) Position of Trap 1 during the mixing process (blue crosses) together with the calculated position of the trap using the velocity and the stop time as fitting parameters (red curve). The trap position was determined from digital images of the coils taken at well defined times during the process. The calculated position only fits the measured one if the stop time is correctly chosen to within about 10 ms.}
\label{trapposition}
\end{figure}

The current ramp must be accurately synchronized with the movement of Trap 1. We apply a reset procedure on the mechanical positioning system to eliminate day-to-day drifts in the trap position. After reset the uncertainty in the coil position is about 50 $\mu$m. By taking images of the trap coils during the movement we have determined the exact position as a function of time (Fig. \ref{trapposition}) and used this information to time the current ramp such that the current in Trap 1 becomes zero exactly when the trap has reached the final position. By using the same procedure we can conclude that the timing drifts of the trap movement is less than 10 ms and this level of stability is indeed necessary to get reproducible results.

\begin{figure*}[!]
\includegraphics[width=\textwidth]{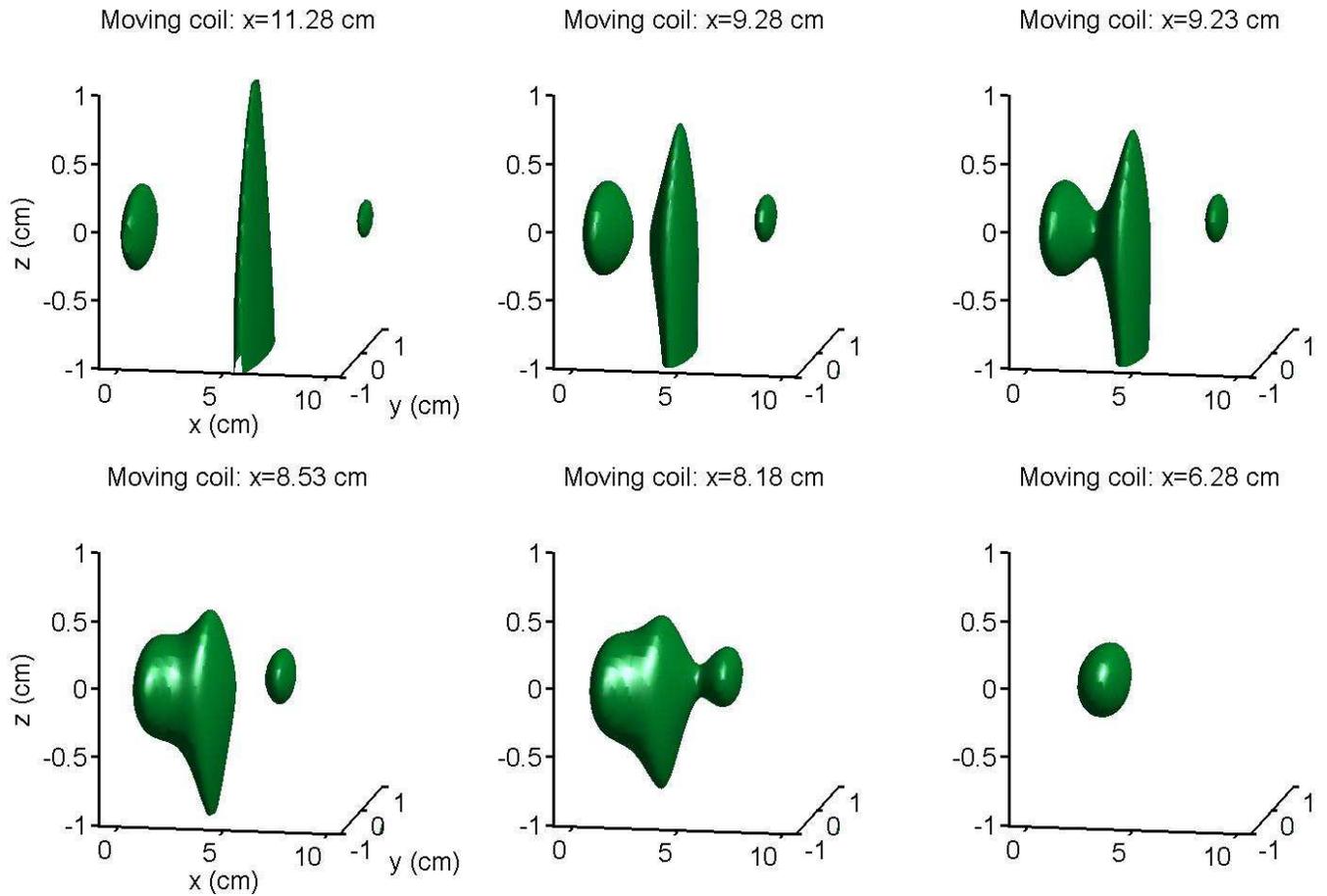}
\caption{\small (Color online) Selection of potential energy isosurfaces during the mixing process with optimal parameters. The surfaces correspond to a potential energy of g$_\textrm{F}m_\textrm{F}\textrm{k}\times 2$ mK. For $F=1$ atoms at a temperature of 200 $\mu$K 65 \% of the atoms would stay inside these isosurfaces in thermal equilibrium (however, we stress that the atoms are not in thermal equilibrium during the mixing process). Notice that the time interval between the third and the fifth plot is only 0.2 s. The vertical asymmetry at large distances from the z=0 plane is due to gravity. The current ramps are as in Fig. \ref{currents}. The current in Trap 2 (situated at $x=0$ cm) is $I_2$=265 A and the initial current in Trap 1 is $I_1=$322 A corresponding to the ratio $I_1/I_2=$1.215.}
\label{isosurf}
\end{figure*}

Armed with a precise magnetic field calculation we are able to study the geometry of the trap potential during the mixing process. This is best viewed by plotting potential energy isosurfaces of the magnetic field. Some representative isosurfaces are shown in Fig. \ref{isosurf}. The shown isosurfaces correspond to an energy of $g_\textrm{F}m_\textrm{F}\textrm{k}\times$ 2 mK where $k$ is Boltzmann's constant. Gravity is also included in the calculations. If the atomic clouds were in thermal equilibrium during the mixing process (and we stress that they are not) 35 \% of the atoms would reside outside these isosurfaces at a temperature of 200 $\mu$K. Therefore it is certainly not clear from these isosurfaces that the mixing can be done successfully, but nevertheless they give an impression of what is going on.

When the two traps are far apart there are 3 isosurfaces - one around the two trap centers and one between the two traps. The isosurface in the middle contains no atoms in the beginning.

When the traps are moving closer together the potential gets weaker and the isosurfaces begin to mix which makes it possible for the atoms to leak into the region in the middle. This is the critical point since the atoms are thermodynamically allowed to spread out over a large volume. That can eventually cause heating when the trap potential tightens again at the end of the mixing process and can lead to loss if atoms hit the wall of the \O\ 40 mm vacuum chamber tube. Shortly after this, all 3 isosurfaces are merged and in the end we are left with only one isosurface which ideally contains all the atoms.

\begin{figure}[!htb]
\includegraphics[width=0.8\columnwidth]{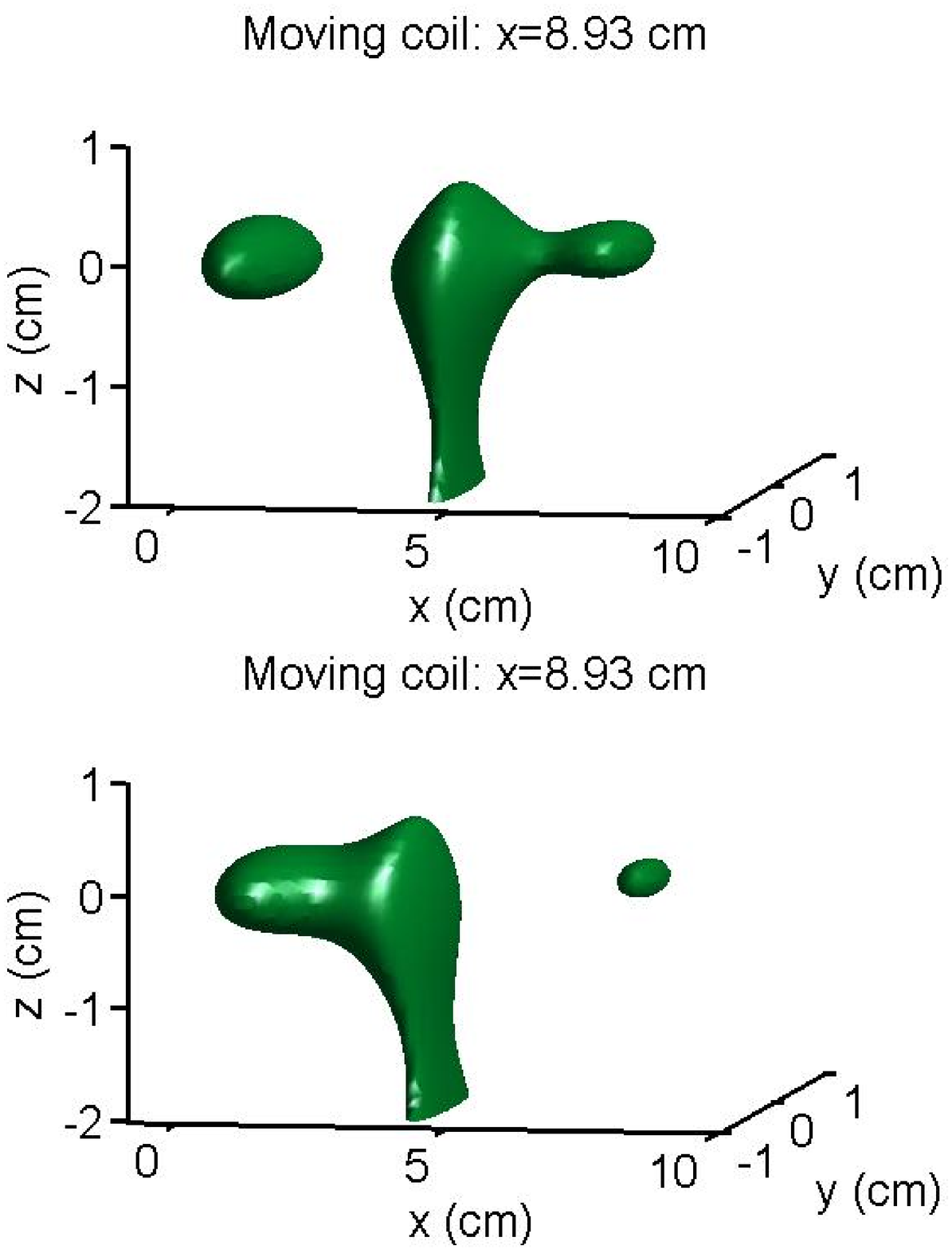}
\caption{\small (Color online) Isosurfaces corresponding to a potential energy of g$_\textrm{F}m_\textrm{F}\textrm{k}\times$ 2 mK for current ratios which are far from optimal. The current in the coils of Trap 2 (situated at $x=0$ cm) is $I_2=$265 A. The initial current in Trap 1 (situated at $x=8.93$ cm) is respectively $I_1=$280 A (upper figure) and 350 A (lower figure). The figure shows that if the current ratio $I_1/I_2$ is too small the atoms in Trap 1 are likely to become lost or heated, if it is too large the atoms in Trap 2 are likely to suffer from loss or heating. Notice that the scales of the axes are different from those in Fig. \ref{isosurf}.}
\label{isosurf2}
\end{figure}

If the current $I_1$ in Trap 1 is too small compared to the current $I_2$ in Trap 2, then the isosurface of Trap 1 mixes with the isosurface in the middle too early and the atoms in this trap are lost. On the other hand, if the ratio $I_1/I_2$ is too large, the atoms in Trap 2 are lost (see Fig. \ref{isosurf2}). This is exactly what we see experimentally, and fortunately there is a well defined current ratio which leads to successful mixing.

\section{Classical simulation of the dynamics}\label{Model}
In order to get a theoretical understanding of the mixing process we have carried out simulations of the particle dynamics. The temperature of the clouds is so high that quantum effects can be safely neglected. One way to see this is to consider the typical spacing between quantum mechanical energy levels in the linear trap potential. From dimensional arguments, or by looking at the energy levels in the 1-dimensional potential $V(x)=C|x|$ (\cite{Sakurai} p. 109) this spacing is of the order of
\begin{align}
E=\left(\frac{C^2\hbar^2}{m}\right)^{1/3}=k\times 0.2\ \mu K
\end{align}
where $k$ is Boltzmann's constant using the worst case parameter $C=\mu_B\times 150$ G/cm. This is 3 orders of magnitude smaller than the typical thermal energy. Furthermore we neglect interactions and collisions between the particles. See section \ref{Thermodynamics} for a discussion of this approximation.

So we propagate the particles classically:
\begin{align}
\mathbf v_i=\frac{\mathrm{d\mathbf r_i}}{\mathrm{dt}}
\qquad m_i\frac{\mathrm{d\mathbf v_i}}{\mathrm{dt}}=-\nabla U(\mathbf r_i)
\end{align}
$\mathbf r_i$ being the position of each particle. We implement $\nabla U=g_\textrm{F}m_\textrm{F}\mu_\textrm{B}\nabla B+mg\hat z$ numerically and solve the equations using a standard multi-order Runge-Kutta method with error control in Matlab. The dynamics is rather complicated and cannot be described analytically. However, all particle trajectories exhibit some oscillatory behaviour with a characteristic period with an order of magnitude given by the cyclotron period
\begin{align}
2\pi\sqrt{\frac{m\rho}{g_\textrm{F}m_\textrm{F}\mu_B\times\partial |B|/\partial\rho}}
\end{align}
which is about 40 ms initially.

\begin{figure}[!]
\includegraphics[width=\columnwidth]{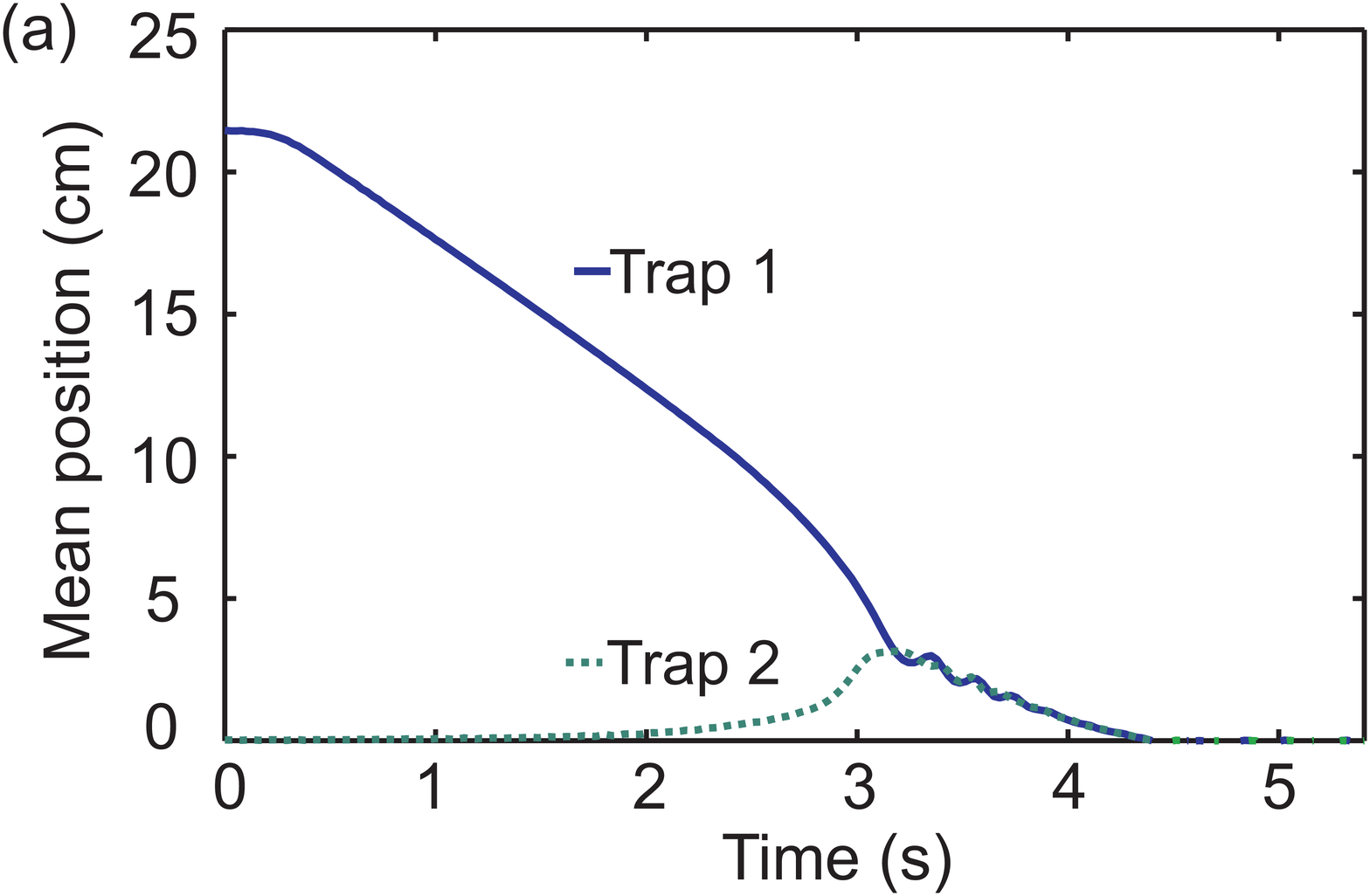}
\includegraphics[width=\columnwidth]{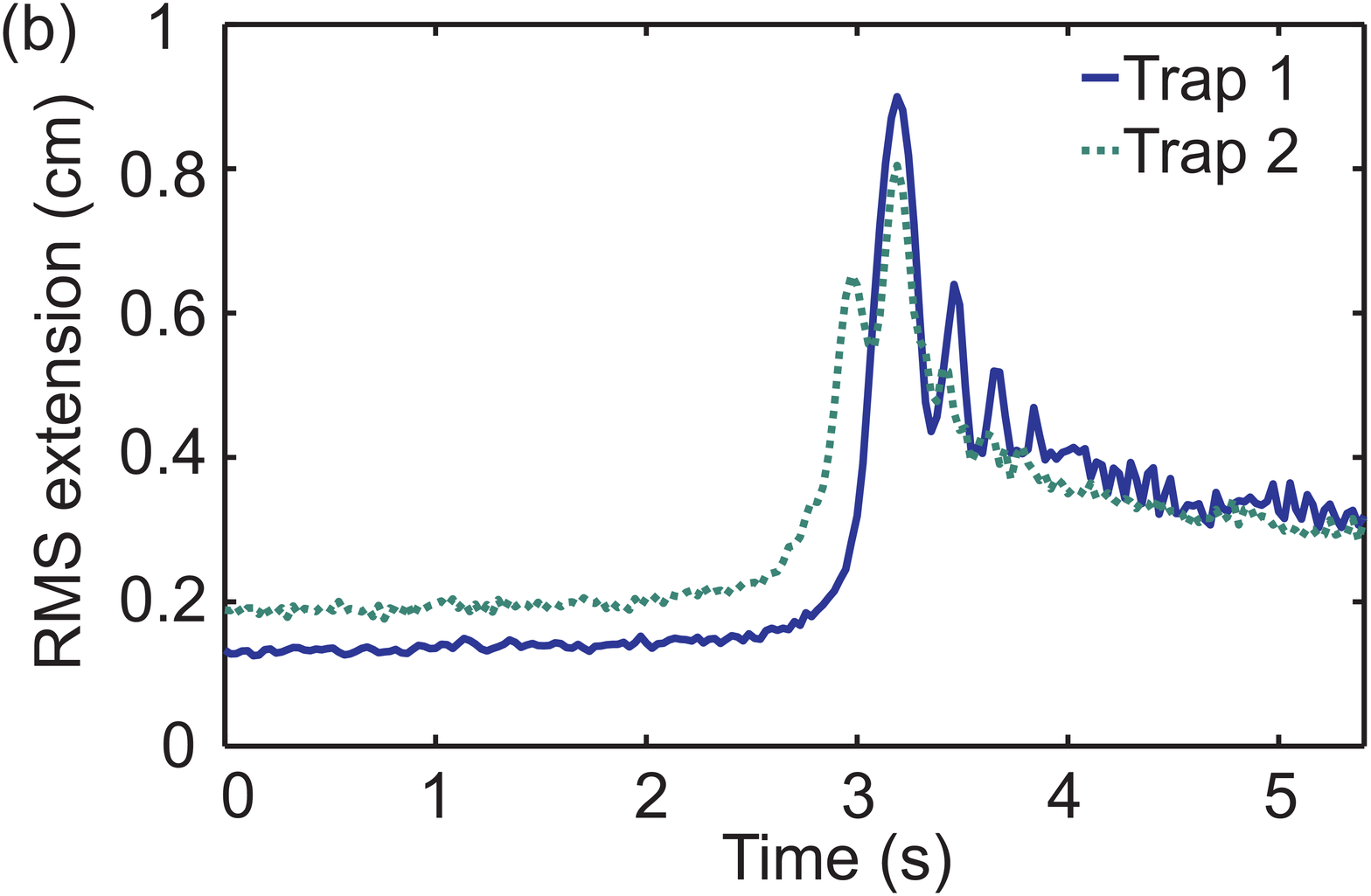}
\\[12pt]
\includegraphics[width=\columnwidth]{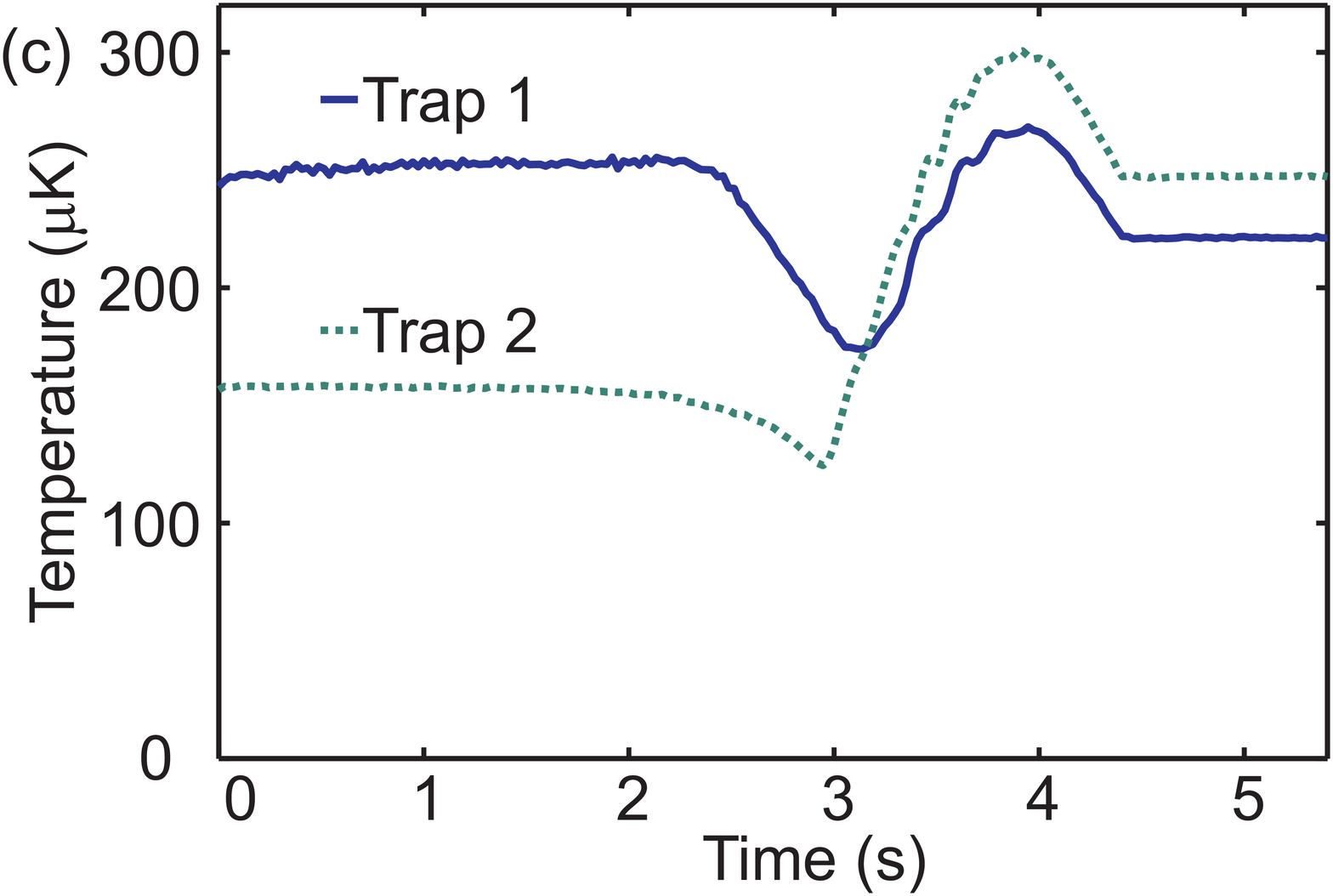}
\caption{\small (Color online) Results of a simulation of the particle dynamics using the current ratio $I_1/I_2=$1.215 which gives the optimal mixing results. (a) shows the position of the two clouds (the mean position of the atoms) as a function of time. (b) shows the RMS extension of the clouds in the mixing direction. (c) shows the mean energy of the two clouds expressed as the corresponding equilibrium temperature $T=(2/9k)<E>$. The temperature temporarily decreases just before the clouds merge, because the trap potential opens up. Since the potential energy gradient of Trap 2 is smaller than the initial gradient of Trap 1 the phase space density of the cloud in Trap 1 decreases even if the final temperature is not higher than the initial one. Particles which are lost during the mixing process have been removed from the calculation.}
\label{PosPlot}
\end{figure}

From the result of such a simulation we can extract various interesting quantities like position, extension and temperature (Fig. \ref{PosPlot}) of the atomic clouds as a function of time. It is easy to discriminate the lost particles from the particles which stay trapped since the former perform a nearly free fall in the field of gravity and will quickly travel far away from the trapping region. Since our vacuum chamber tube has a diameter of 4 cm, particles which during the process reach a position which is more than 2 cm away from the transport axis are considered lost and removed from the calculation before calculating the quantities above.

Using optimal parameters it turns out that if we consider the atoms originating from Trap 1 and Trap 2 individually, then the kinetic and potential energies correspond to the same effective temperature and there is almost no anisotropy of the velocity distribution. This means that these ensembles are not far from thermal equilibrium at the end of the mixing process. But the ensemble from Trap 1 is not in thermal equilibrium with the ensemble from Trap 2. However, since thermalization takes place through elastic collisions the equilibrium temperature of the mixed cloud as a whole can be found from the conserved mean energy, since in thermal equilibrium $<E>=<E_\textrm{kin}>+<U>=(9/2)kT$ as described in section \ref{Thermodynamics}. Also the equilibrium phase space density can be calculated.

\subsection{Thermodynamics of the quadrupole trap}\label{Thermodynamics}
Using the approximation (\ref{fieldapprox}) thermodynamic equilibrium quantities of the trapped atoms can be calculated. We introduce the radial trap potential gradient
\begin{align}
C=g_\textrm{F}m_\textrm{F}\mu_B\frac{\partial |B|}{\partial\rho}
\end{align}
The density is given by the Boltzmann distribution:
\begin{align}
n(\mathbf r)&=\frac{N}{4\pi}\left(\frac{C}{kT}\right)^3e^{-C\sqrt{\rho^2+4z^2}/kT}
\end{align}
where $N$ is the number of atoms.

The peak phase space density is
\begin{align}\label{PSDformula}
n(\mathbf r=\mathbf 0)\lambda_\textrm{dB}^3&=\frac{N}{4\pi}\left(\frac{C}{kT}\right)^3\left(\frac{h^2}{2\pi mkT}\right)^{3/2}\propto \frac{NC^3}{T^{9/2}}.
\end{align}

The mean kinetic and potential energies are
\begin{align}\label{Eformula}
<E_\textrm{kin}>=\frac{3}{2}kT
\qquad
<U>=3kT
\end{align}
arising solely from the fact that the energy is quadratic in momentum and linear in position.

In order to determine the mean elastic collision rate we need to know the mean density which is given by
\begin{align}
<n(\mathbf r)>&=\int d^3\mathbf r\ n(\mathbf r)\frac{n(\mathbf r)}{N}=\frac{1}{8}n(\mathbf r=0)
\end{align}
The mean time between elastic collisions becomes
\begin{align}\label{CollRate}
\tau_\textrm{el}&=(<n>\sigma <v_\textrm{rel}>)^{-1}
\\&=19\textrm{ ms}\cdot\frac{10^8}{N}\left(\frac{T}{100\ \mu\textrm{K}}\right)^{5/2}\left(\frac{1}{g_\textrm{F}m_\textrm{F}}\frac{100\textrm{ G/cm}}{\partial |B|/\partial\rho}\right)^3
\end{align}
using the scattering length $a=106\ a_0$ from \cite{Sympathetic8587,Burke}, the cross section $\sigma=8\pi a^2$ and the mean relative velocity $<v_\textrm{rel}>=\sqrt{8kT/(\pi m/2)}$. For the experimental parameters used in this work $\tau$ lies in the range 1-8 s in the science chamber trap. This means that thermalization is expected to occur after the mixing process has taken place. Experimentally our coldest clouds (those we get when we use optimal parameters) are thermalized shortly after the transfer to the science chamber trap while the hotter clouds take on the order of 5-30 s to thermalize.

During the mixing process, where the magnetic fields of the two traps interfere, the mean density is smaller so the mean time between collisions is even larger. At a mixing speed of 5 cm/s the mixing effectively takes place in a fraction of a second, which implies that thermalization is highly unlikely to be important. Therefore it is reasonable to neglect collisions in the calculations during the mixing process. After the mixing process, however, the clouds begin to thermalize and experimentally we store the atoms for 10 s in the science chamber quadrupole trap before we take the image in order to be close to thermal equilibrium.

\subsection{Initial conditions}
In each trap we prepare an ensemble of 500 particles obeying Boltzmann statistics. We assume that the phase space density is conserved during  transport and adiabatic compression \cite{Ketterle} so we choose the initial temperature such that the phase space density matches the one determined from temperature measurements in the science chamber. At a radial trap gradient $\partial |B|/\partial\rho$ of 84 G/cm we measure the temperature of a cloud which has not gone through the mixing process to be 195 $\mu$K. Using the scaling of the phase space density (\ref{PSDformula}) we see that the proper initial temperature is
\begin{align}
T_\textrm{init}=\left(\frac{84\textrm{ G/cm}}{\partial |B|/\partial\rho}\right)^{-2/3}\times 195\ \mu\textrm{K}
\end{align}
so the two clouds have different temperatures but the same phase space density.

The initial velocities are found by selecting each Cartesian component randomly from the Gaussian distribution:
\begin{align}
P(v_j)\propto e^{-\frac{1}{2}mv_j^2/kT_\textrm{init}},\quad j=x,y,z
\end{align}
We distribute the positions using a rejection method, see Footnote \cite{rejection}.

\section{Results}\label{Results}
Using optimal parameters we get the following main result: If we mix two clouds with about $1.7\times 10^{8}$ atoms each, we loose only about 5\% of the atoms and the measured temperature in the science chamber increases from 195 $\mu$K to 310 $\mu$K. This means that according to Eq. (\ref{PSDformula}) the phase space density relative to the phase space density of the initial clouds is 24 \%. Using optimal parameters in the theoretical model we get a relative phase space density of 27 \% in excellent agreement with the experimental result. The elastic collision rate is also important in relation to subsequent evaporative cooling. It scales as $N/T^{5/2}$ (see Eq. (\ref{CollRate})) which means that the mixed cloud has an elastic collision rate of about 60 \% of the rate of the initial clouds. Based on this we conclude that it should be possible to cool the mixed cloud to quantum degeneracy using evaporative cooling.

The most critical parameter of the mixing process is the ratio $I_1/I_2$ of the currents in the coils of Trap 1 and Trap 2. This is discussed in section \ref{VaryRatio}. There are several other parameters which can be varied in order to optimize the process. Here we look at the trap depth (section \ref{TrapDepth}) and the initial temperature (\ref{InitialTemp}).

\subsection{Dependence on trap current ratio}\label{VaryRatio}
\begin{figure}[!]
\includegraphics[width=\columnwidth]{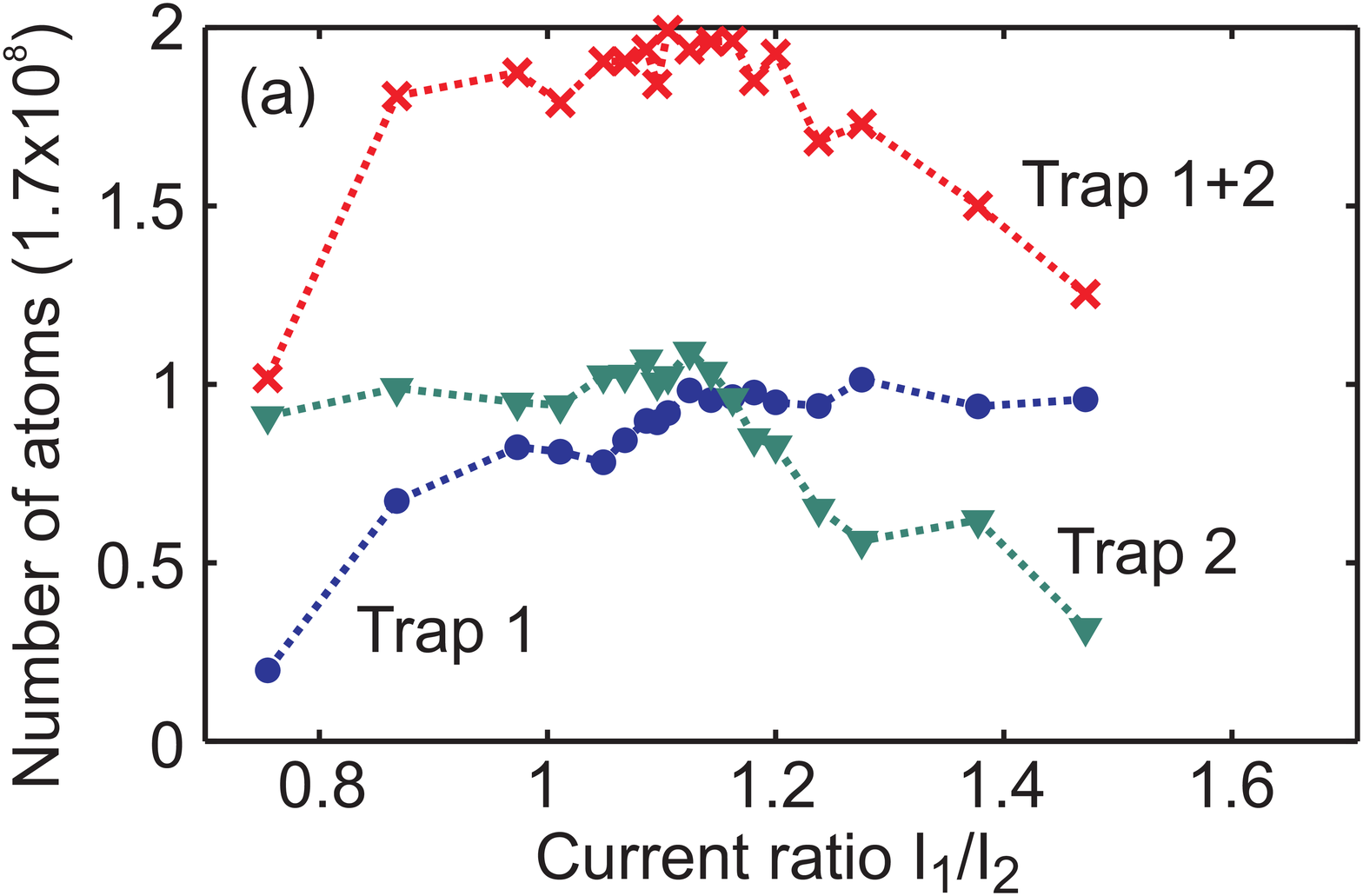}
\\[12pt]
\includegraphics[width=\columnwidth]{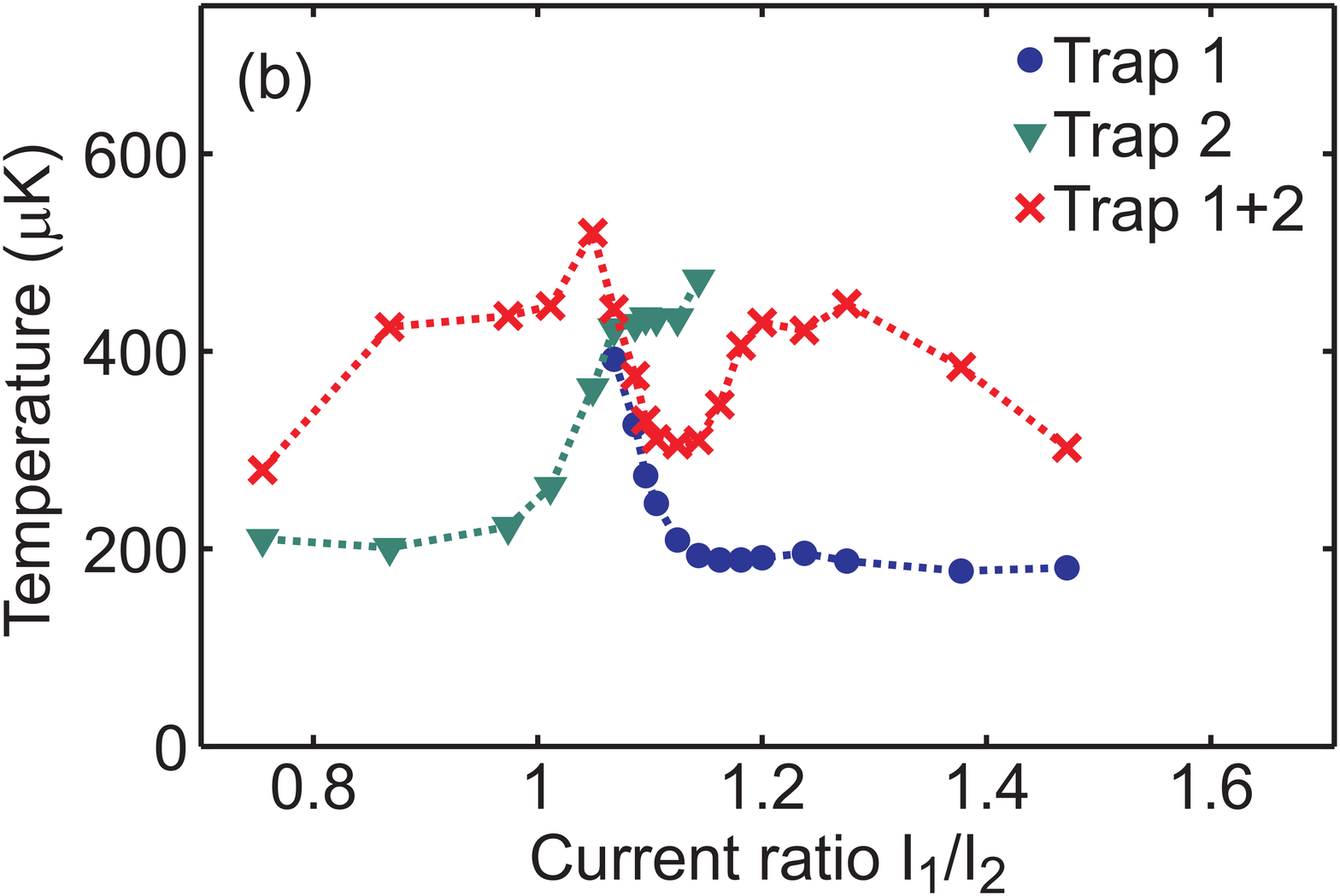}
\includegraphics[width=\columnwidth]{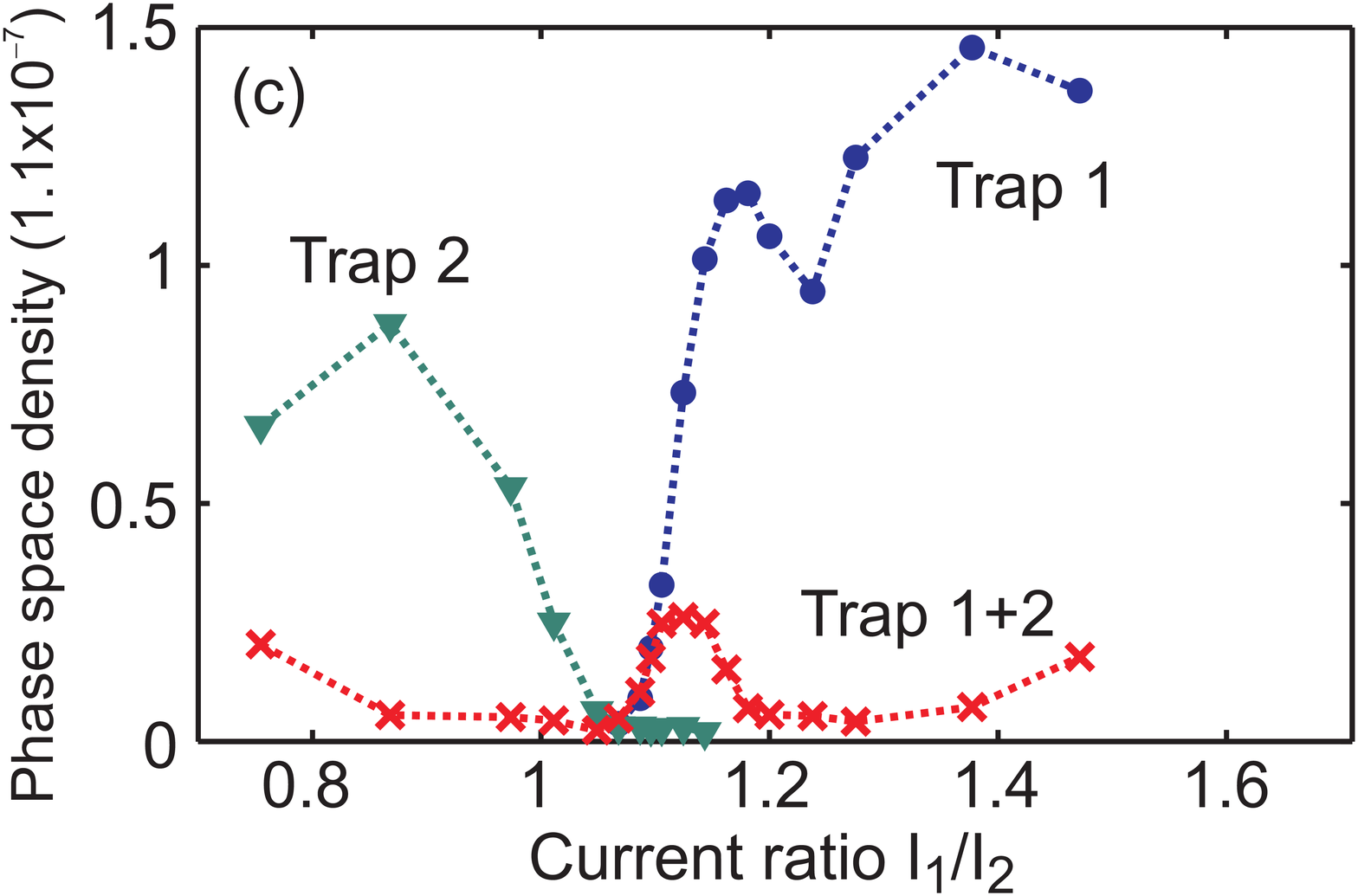}
\caption{\small (Color online) Experimental results of varying the ratio $I_1/I_2$ of the currents in the coils of Trap 1 and Trap 2, respectively ($I_2$ is kept constant while $I_1$ is varied). (a) shows the number of atoms staying trapped during the process (normalized to the number of atoms in each of the initial clouds which is $1.7\times 10^8$), (b) shows the final temperature and (c) shows the final phase space density (normalized to the value $1.1\times 10^{-7}$ which is the phase space density of an initial cloud with $1.7\times 10^8$ atoms and a temperature of 195 $\mu$K). All three plots contain three curves: One curve showing results with atoms in both traps and two curves showing results with atoms in only one or the other of the traps. The dotted lines between the points are drawn to guide the eye.}
\label{ExpRatio}
\end{figure}

\begin{figure}[!]
\includegraphics[width=\columnwidth]{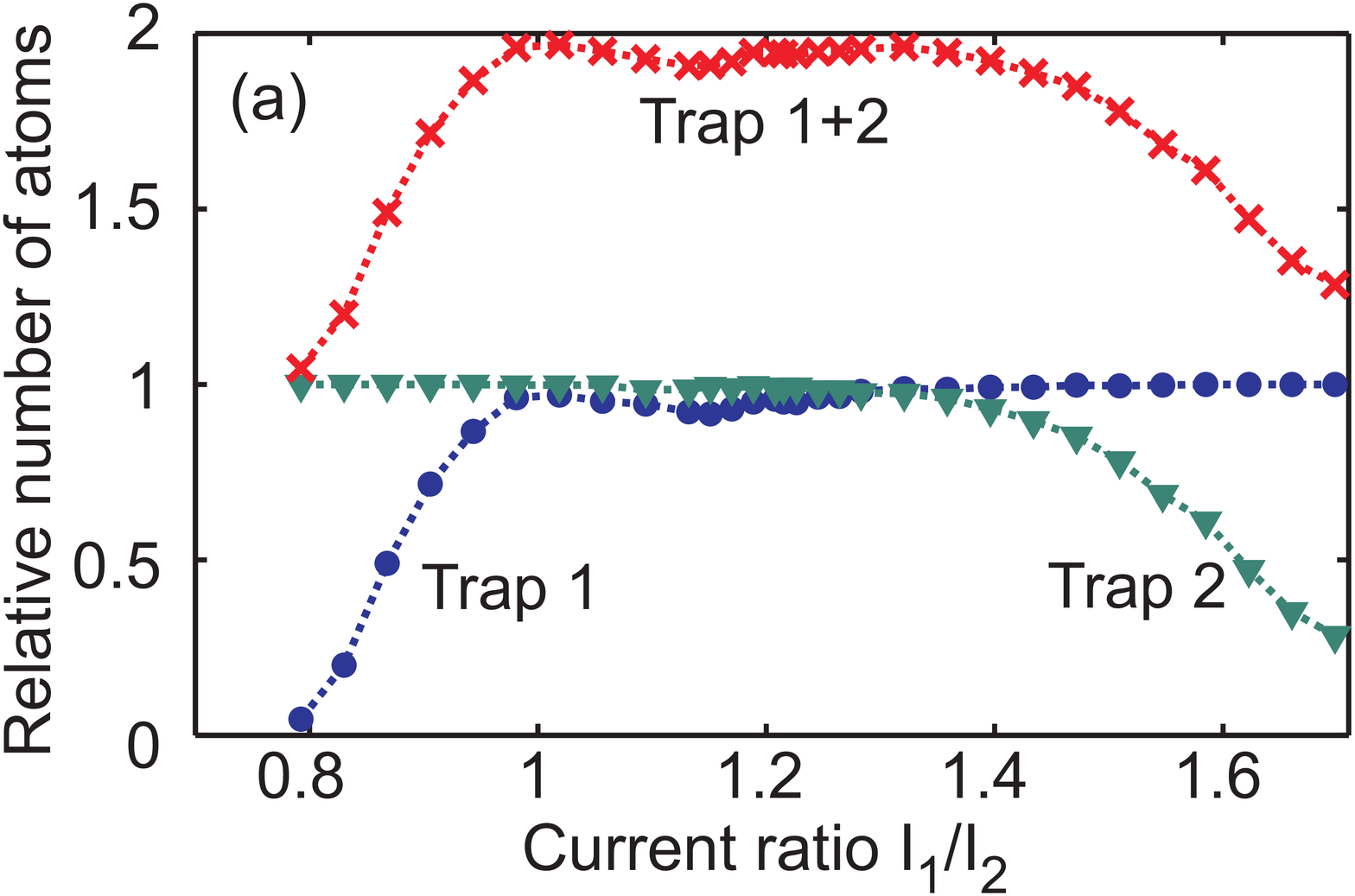}
\\[12pt]
\includegraphics[width=\columnwidth]{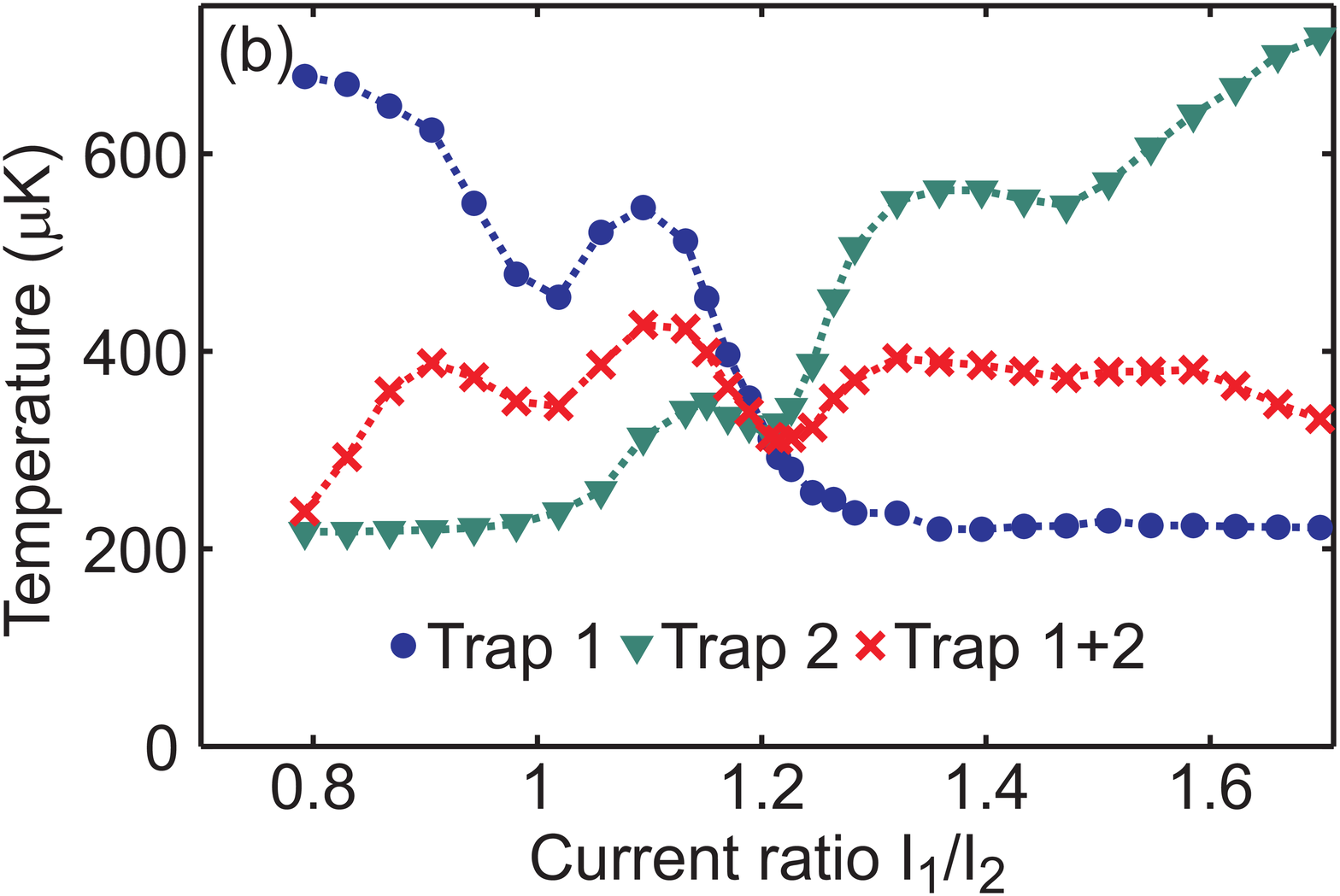}
\includegraphics[width=\columnwidth]{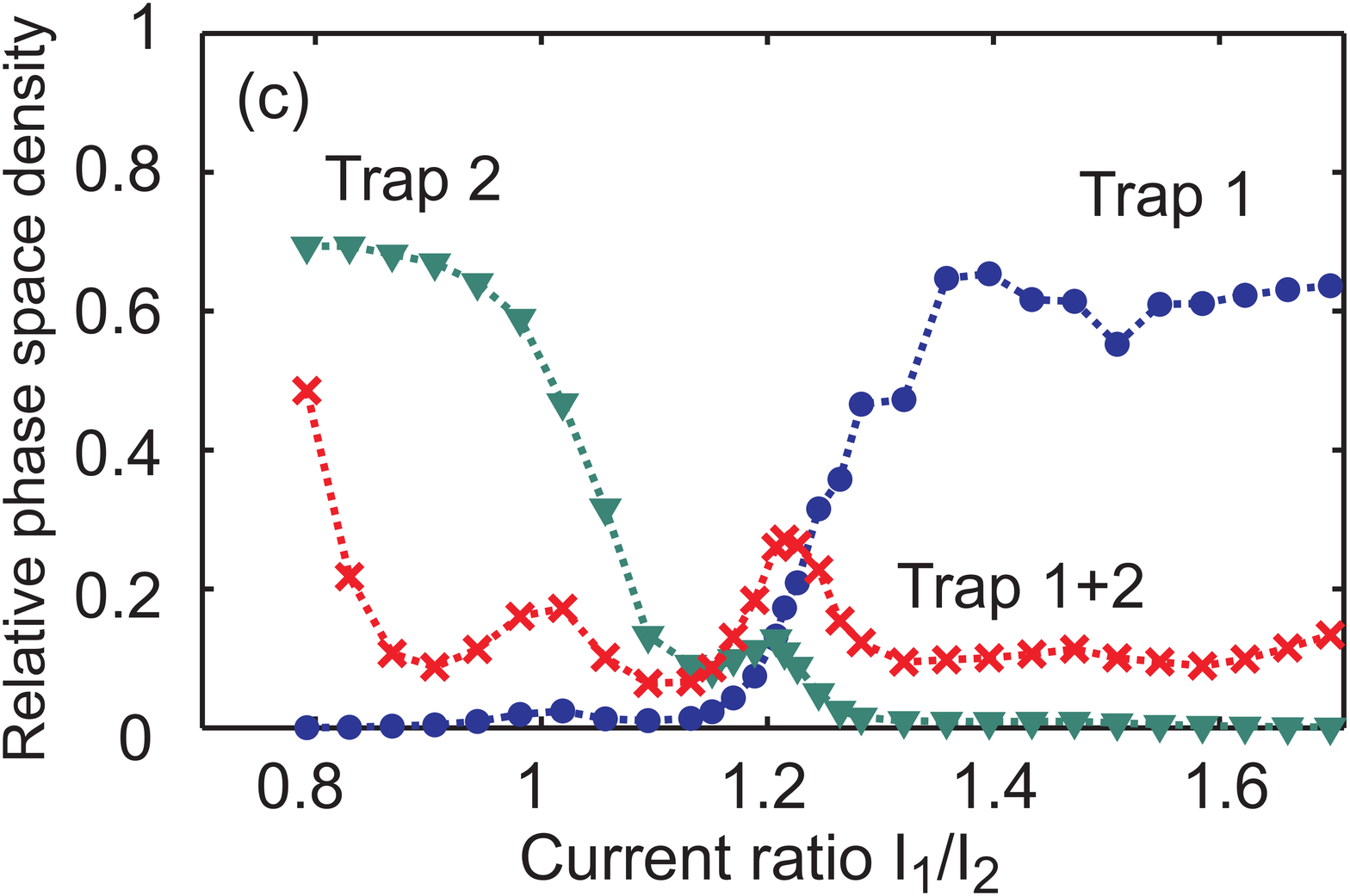}
\caption{\small (Color online) Theoretical results of varying the current ratio. The plots show the same quantities as the plot in Fig. \ref{ExpRatio}. Atoms which escape the trap potential or hit the chamber wall during the mixing process are removed before calculating the temperatures and phase space densities. The number of atoms and the phase space density are normalized to the (identical) values of each of the initial clouds.}
\label{TheoryRatio}
\end{figure}

Keeping $I_2$ constant at 265 A which corresponds to a radial magnetic field gradient of 53 G/cm and changing the current ratio by varying $I_1$, we get the experimental results shown in Fig. \ref{ExpRatio}. The corresponding results of the theoretical simulations with the same parameters lead to the results in Fig. \ref{TheoryRatio}. In these figures we plot (a) the number of atoms staying trapped during the process, (b) the final temperature, and (c) the final phase space density. In each of these plots the results are given for atoms in both traps and with atoms in only one or the other of the two traps. Thereby it is possible to determine which of the two clouds is lost or heated most during the process. It should be stressed that the uncertainty of the phase space density value is relatively large, because it decreases rapidly with temperature (see Eq. \ref{PSDformula}). A change in phase space density by a factor of 2 corresponds only to a relative temperature change of 17 \%.

Both the experimental and theoretical plots show that for large current ratios the atoms in Trap 1 stay trapped during the mixing process without loss or heating, whereas a small current ratio favors the atoms in Trap 2. This confirms what one could expect from the isosurface plots in Fig. \ref{isosurf2}. Experimentally we see that when we have atoms in Trap 1 only and use a current ratio less than 1.06, we get a diffuse cloud which is very elongated in the horizontal direction and not in thermal equilibrium. Therefore we cannot attribute a temperature to these clouds \cite{EvenIf}. This is the reason why the "Trap 1" curve stops in the temperature and phase space density plots. The same situation applies when we have atoms only in Trap 2 at a current ratio exceeding 1.14. The number of atoms in these cases is determined from a sum of pixel values after 2 ms time-of-flight.

Fortunately it is possible to choose a current ratio which constitutes a compromise where the heating of the mixed cloud as a whole is small and almost no atoms are lost. Both experimentally and theoretically there is a well defined current ratio which at the same time minimizes the temperature and maximizes the number of atoms of the mixed cloud. We will refer to this current ratio as the optimal current ratio. Experimentally it is 1.125 while in the theoretical calculations it is 1.215. We believe that the 8 \% difference between these two values can be attributed to uncertainties in the calculation of the magnetic field from the coils.

The qualitative agreement between the measurements and the theoretical simulations is striking and also, bearing in mind that the optimal current ratio is slightly different, even the quantitative agreement is not bad. Similar to the experimental results, the theoretical simulations give a final phase space density of about 27 \% of the initial phase space density of the initial clouds at the optimal current ratio.

But there are also differences. The simulations result in two local maxima of the final phase space density whereas experimentally only one clear maximum is observed. Also, theoretically at the optimal current ratio the phase space density of the atoms from each of the two traps is almost equal. Experimentally the phase space density is significantly larger when we have atoms only in Trap 1 than when we have atoms only in Trap 2 at the optimal current ratio. Taking the simplicity of the theoretical model into account one cannot expect perfect  quantitative agreement. We recall that the model does not include the dynamics before and after the mixing process and it ignores collisions and interactions.

\subsection{Dependence on trap depth}\label{TrapDepth}
To the extent that gravity can be neglected, changing the trap depth only changes the magnitude and not the geometrical shape of the potential. Therefore this parameter is not expected to be very critical as long as the traps are deep enough that the atoms do not escape. This is confirmed by our measurements. First of all there is no measurable change in the optimal current ratio when we change the trap depth. For small trap depths the final temperature has no well defined minimum as we vary the current ratio, but the number of atoms peaks at a current ratio which is independent of trap depth. Therefore it makes sense to vary the trap depth keeping a constant current ratio.

\begin{figure}[!]
\includegraphics[width=\columnwidth]{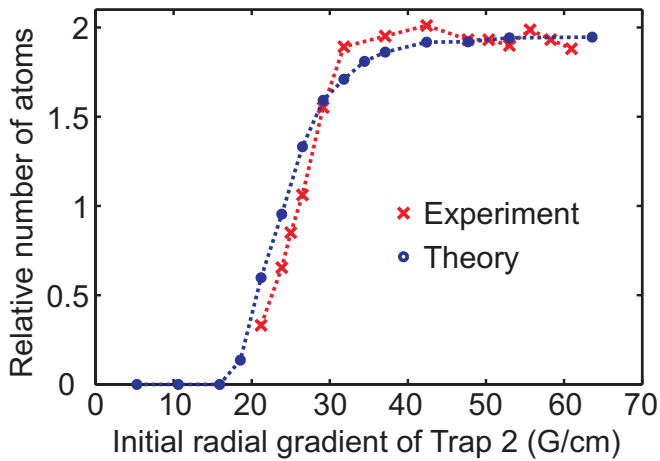}
\caption{\small (Color online) The number of atoms staying trapped during the mixing process relative to the initial number of atoms in each trap for different trap gradients. The measurements are normalized to 1.81$\times 10^8$ atoms. The current ratio is $I_1/I_2=1.125$ for the experimental results and 1.215 for the theoretical results. As long as the initial radial field gradient of Trap 2 is larger than 35 G/cm (corresponding to an initial barrier height of 92 G in Trap 2) the loss of atoms is only about 5 \%. When the field gradient becomes smaller than this value a significant loss of atoms sets in. The experimental and theoretical results agree very well.}
\label{TrapDepthSurvival}
\end{figure}

The number of atoms remaining trapped after mixing versus the trap depth is shown in Fig. \ref{TrapDepthSurvival}, where the experimental and theoretical results are compared. They agree very well. For sufficiently deep traps there is essentially no loss of atoms. When the trap gradient becomes smaller than a critical value which is about 60 \% of the value used above (which corresponds to a current of 265 A in Trap 2 giving an initial radial field gradient of 53 G/cm and a barrier height of 154 G - see Table \ref{coilpar}) then trap losses start to set in. At a trap depth of 40 \% of the value used above only 14 \% of the atoms are left. We have also measured how the temperature of the mixed cloud depends on the trap depth. In the region with essentially no loss of atoms the temperature variations are less than 10 \%. It should be kept in mind that these results are obtained for the $|F=1,\ m_\textrm F=-1>$ hyperfine state which has a magnetic moment of $\mu_\textrm{B}/2$. Using the $|F=2,\ m_\textrm F=2>$ state, which has a magnetic moment of $\mu_\textrm{B}$, the same potential energy gradient can be achieved with a field gradient which is only half as large.

\subsection{Theoretical dependence on the initial temperature}\label{InitialTemp}
\begin{figure}[!]
\includegraphics[width=\columnwidth]{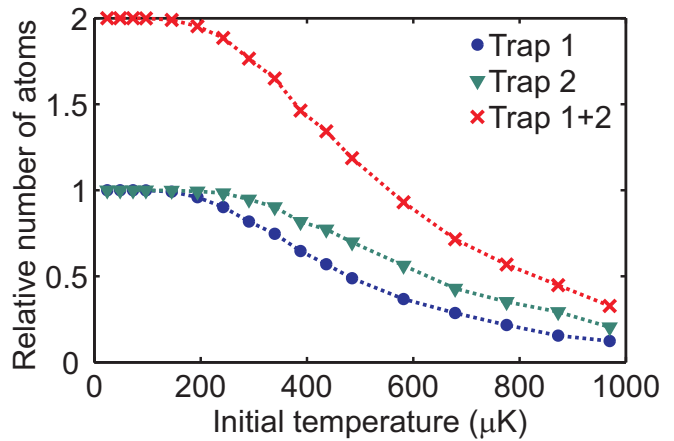}
\caption{\small (Color online) Results of simulations showing the fraction of atoms staying trapped during the mixing process for different initial temperatures. For the trap gradient used (53 G/cm in trap 2) there is essentially no loss of atoms until the initial temperature rises above about 200 $\mu$K. At a temperature of 1000 $\mu$K only 15 \% of the atoms stay trapped. The current ratio is 1.215. Each point corresponds to a simulation using 2$\times$2000 atoms initially.}
\label{TheoryTempSurvival}
\end{figure}

\begin{figure}[!]
\includegraphics[width=\columnwidth]{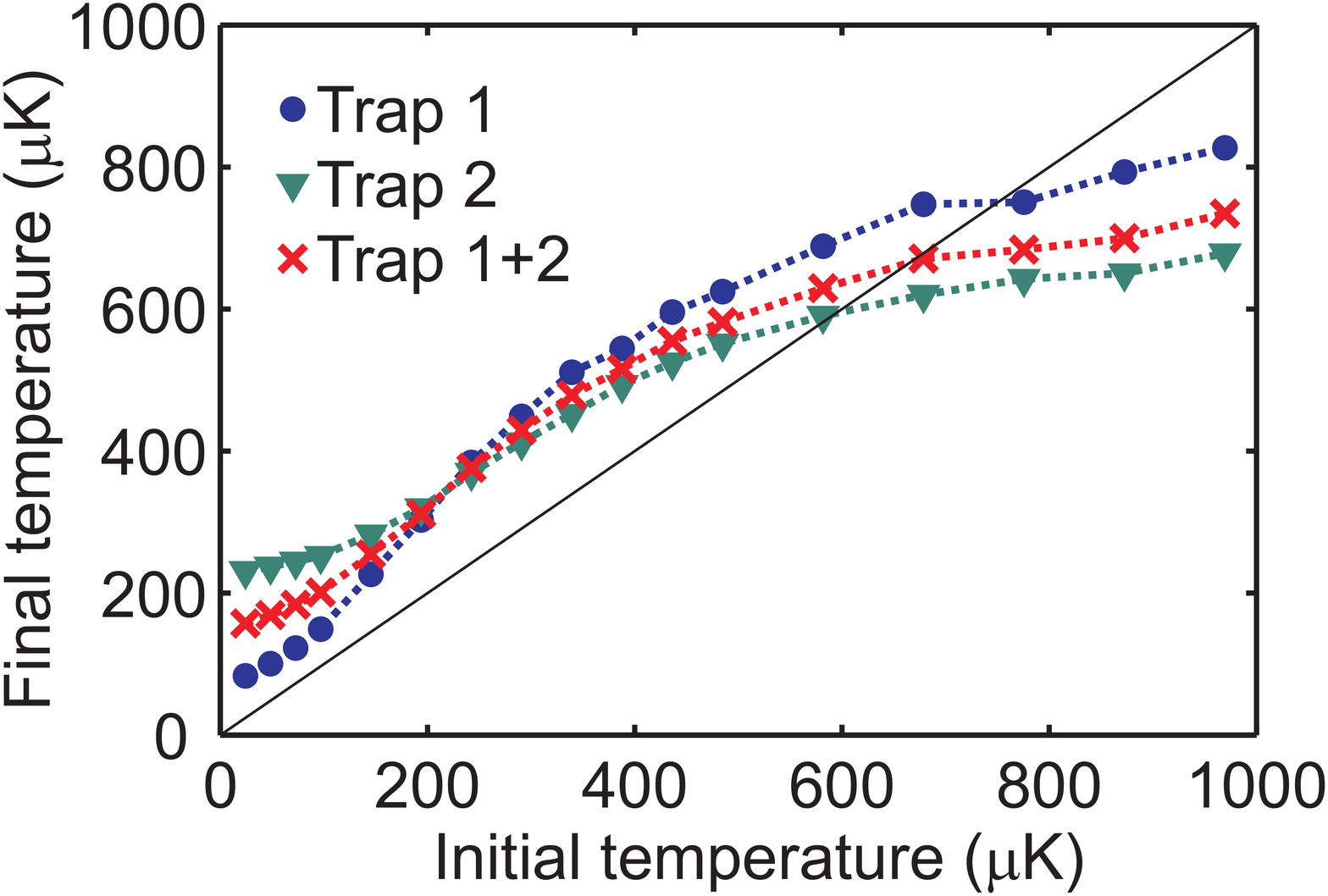}
\caption{\small (Color online) Theoretical dependence of the temperature after the mixing process on the initial temperature at a current ratio of 1.215 which is optimal for low temperatures. Up to an initial temperature of about 500 $\mu$K the absolute temperature increase is roughly independent of the initial temperature. For very low initial temperatures, the final temperature converges towards 150 $\mu$K for a mixture with equal numbers of atoms in the two traps. At this stage the properties of the mixed cloud are completely determined by the mixing process and independent of the initial temperature. Apparently for the trap gradients used the process is also feasible for higher temperatures, but then the most energetic atoms escape or hit the vacuum chamber wall. The thin black line shows the initial temperature for comparison.}
\label{TheoryTempTemp}
\end{figure}

To check how generally applicable the mixing procedure is, it is also interesting to look at the dependence on the initial temperature of the clouds which are mixed. We have studied this in the theoretical simulations using the radial field gradient 53 G/cm in Trap 2. In all the calculations we have used the current ratio 1.215 which is theoretically optimal at low temperatures (as long as there is essentially no loss of atoms). For higher temperatures the optimal current ratio is less well defined, because there is no clear minimum of the final temperature. This is because there is a significant loss of atoms and the most energetic atoms have the largest probability of escaping. The fraction of atoms remaining trapped after the mixing process is plotted in Fig. \ref{TheoryTempSurvival} and the final temperature is plotted in Fig. \ref{TheoryTempTemp}.

The figures show that up to an initial temperature of 200 $\mu$K all atoms stay trapped whereas for higher temperatures the loss gradually increases until at 1 mK 85 \% of the atoms are lost. On the other hand one does not gain very much from having an even colder cloud initially since the absolute temperature increase is about 100 $\mu$K and roughly independent of the initial temperature up to an initial temperature of about 500 $\mu$K. For even higher temperatures the final temperature drops below the initial temperature, because the most energetic atoms escape the trapping potential or hit the vacuum chamber wall.

\section{Outlook: Different species}\label{Outlook}
Although of limited practical interest, mixing of two Rb clouds is an excellent way to study the process of mixing atomic clouds by merging two quadrupole traps. The perspective is to use the process to mix two different species. We hope that a cold mixture produced in this way can be a good starting point for making a quantum gas mixture. It is our intention to mix $^{87}$Rb and $^6$Li or $^7$Li and do evaporative cooling on Rb to get both species into the quantum regime.

For alkali atoms it is normally preferable to have both species in the stretched hyperfine state with $g_\textrm{F}m_\textrm{F}=1$. Besides having the largest possible magnetic moment, the trapped atoms in such a mixture are least affected by loss due to spin-exchange collisions. Thus, assuming that both species are in a state with $g_\textrm{F}m_\textrm{F}=1$, the only difference between the two species affecting the dynamics is their initial temperature (which is typically the same within a factor of two or so) and their mass which enters via Newton's 2$^\textrm{nd}$ law and via the initial velocity which has the mean value $v=\sqrt{8kT/\pi m}$.

Using our theoretical model which simulates the essential features of the Rb-Rb mixing very well, we can make predictions on the outcome of an experiment where two different species are mixed and the results are very promising. We have simulated mixing of Rb in Trap 1 with Li, Na, K and Cs in Trap 2. The radial field gradient of Trap 2 was 53 G/cm as above such that the magnetic trapping potential is twice as large as the  potential for the $|F=1,\ m_\textrm F=-1>$ $^{87}$Rb atoms used above. We have chosen the initial temperature 250 $\mu$K at a radial trap gradient of 42 G/cm except for Li for which 500 $\mu$K is more realistic \cite{GrimmLiMOT}.

\begin{figure}[!]
\includegraphics[width=\columnwidth]{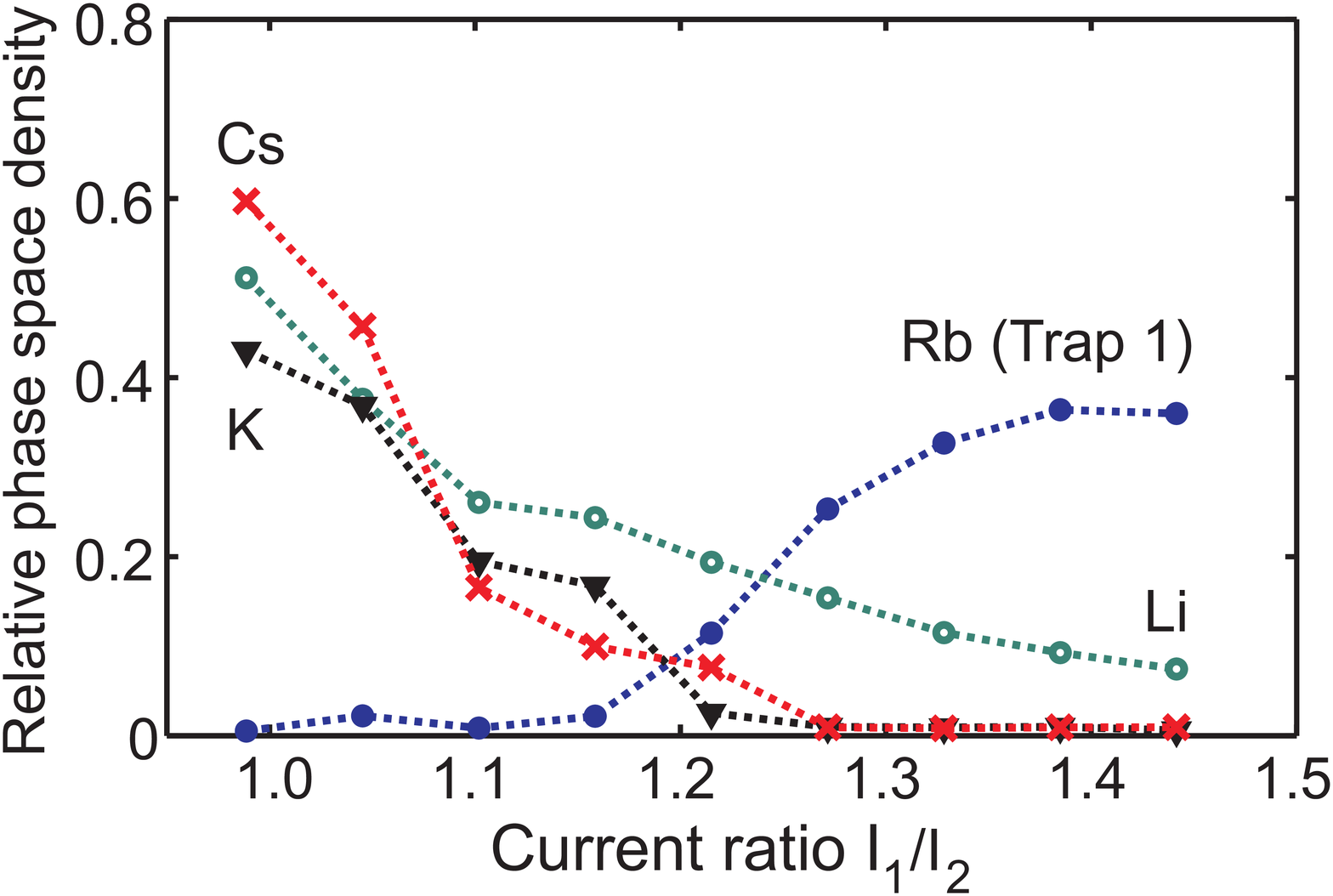}
\caption{\small (Color online) Simulation results of mixing of two different species. The plot shows the final phase space density of each of the clouds resulting from mixing of Rb in Trap 1 ($\bullet$) with respectively Li ($\circ$), K ($\blacktriangledown$) and Cs ($\times$) in Trap 2. The results for Na resembles the results for K very much. The temperature of the initial clouds at a radial field gradient of 42 G/cm is 250 $\mu$K except for Li for which it is 500 $\mu$K. The phase space densities are normalized to the values of the initial clouds. Each point corresponds to a simulation using 2$\times$5000 atoms initially.}
\label{DiffPSD}
\end{figure}

For current ratios in the range 1-1.45 the loss of atoms is below 5 \% except for the Li cloud for which the loss is 8-12 \%. Fig. \ref{DiffPSD} shows the final phase space density relative to the initial phase space density of each of the clouds. Just as for Rb-Rb mixing the current ratio can be adjusted to favor one or the other of the two atomic clouds and there is a region of current ratios where the phase space densities of both species are retained at a reasonable level.

Since evaporative cooling is notoriously a very lossy process, if one of the species is to be sympathetically cooled by the other, one typically wants to have many atoms acting as the cooling agent and fewer atoms to be sympathetically cooled (and it does not matter if the temperature of these few atoms is larger in the beginning as long as their contribution to the total energy is small). Therefore one might benefit from choosing a current ratio which minimizes the loss and heating of the cooling agent while keeping the loss and temperature increase of the other species at an acceptable level.

\section{Conclusion}\label{Conclusion}
In conclusion, we have demonstrated a new method to make mixtures of magnetically trapped, ultracold atoms based on merging of simple magnetic quadrupole traps mounted on industrial translation stages. This method should be generally applicable provided that the trap gradients and the diameter of the vacuum chamber tube in the mixing region are large enough.

Our experimental results show that two clouds of Rb atoms can be mixed with only about 5 \% loss of atoms, and with a final phase space density which is 24 \% of the initial one. This loss of phase space density is so small that it should be possible to cool the mixed cloud down to quantum degeneracy. 

The experimental results are in good agreement with simple classical simulations of the mixing process. Extending these simulations to mixtures of different elements, we get promising results for mixing of different species and we believe that the method will become a useful new way of producing mixtures of ultracold atoms.

\acknowledgments{The authors would like to thank the Faculty of Science at the University of Aarhus, the Danish National Research Foundation Center for Quantum Optics (QUANTOP), the Carlsberg foundation, and the Danish Natural Science Research Council for financial support. We also thank the local staff for invaluable technical support. Finally we thank Klaus M\o lmer for inspiring discussions and for useful comments to the manuscript.}


\begin{thebibliography}{99}
\bibitem{rejection} First we choose random coordinates uniformly (within intervals so large that the probability for being outside this volume is less than $10^{-4}$). The probability that the chosen position occurs physically is proportional to $e^{-U(\mathbf r)/kT_\textrm{init}}$ so by choosing another random number $p$ uniformly between 0 and 1 we can decide whether this position shall be accepted (if $p<e^{-U(\mathbf r)/kT_\textrm{init}}$) or rejected (if $p>e^{-U(\mathbf r)/kT_\textrm{init}}$).
\bibitem{EvenIf} Even when we let the atoms equilibrate for 30-50 s in the science chamber we cannot measure a consistent temperature. The measured temperature seems to increase in time, and from measurements of the more regular clouds we know that this apparent temperature increase is not due to technical noise.
\end{thebibliography}
\end{document}